\pdfoutput=1
\documentclass[11pt]{article}

\usepackage[final]{acl}

\usepackage{float}
\usepackage{placeins}

\usepackage{times}
\usepackage{latexsym}
\usepackage[most]{tcolorbox}
\usepackage{listings}

\usepackage{longtable}

\usepackage[T1]{fontenc}
\usepackage[utf8]{inputenc}

\usepackage{microtype}

\usepackage{inconsolata}

\usepackage{graphicx}

\usepackage{multirow}
\usepackage{booktabs}
\usepackage{xcolor}

\newcommand{\lqquote}{\textquotedblleft}   % left quote
\newcommand{\rqquote}{\textquotedblright}    % right quote

\lstdefinelanguage{json}{
  basicstyle=\ttfamily\footnotesize,
  breaklines=true,
  frame=single,
  backgroundcolor=\color{gray!10},
  escapeinside={(*@}{@*)}, % Anything between (*@ and @*) is processed as LaTeX code
  literate=
    {:}{{\textcolor{blue}{:}}}{1}
    {,}{{\textcolor{blue}{,}}}{1}
    {[}{{\textcolor{blue}{[}}}{1}
    {]}{{\textcolor{blue}{]}}}{1}
    {\{}{{\textcolor{blue}{\{}}}{1}
    {\}}{{\textcolor{blue}{\}}}}{1}
}

\title{SIFT-50M: A Large-Scale Multilingual Dataset for \\ Speech Instruction Fine-Tuning}

\author{%
{\bfseries
Prabhat Pandey\thanks{Equal contribution.}$^{1}$\quad Rupak Vignesh Swaminathan\footnotemark[1]$^{1}$\quad K V Vijay Girish\footnotemark[1]$^{1}$
}\\[0.5ex]
{\bfseries
Arunasish Sen$^{1}$\quad Jian Xie\thanks{Work done while at Amazon.}$^{2}$\quad Grant P. Strimel$^{1}$\quad Andreas Schwarz$^{1}$
}\\[1ex]
  $^{1}$Amazon AGI \quad
  $^{2}$Apple Inc.\\[1ex]
  Correspondence: \texttt{\{panprabh, swarupak, kvvijayg\}@amazon.com}
}

\begin{document}
\maketitle
\pagestyle{plain}

\begin{abstract}
We introduce SIFT (\textbf{S}peech \textbf{I}nstruction \textbf{F}ine-\textbf{T}uning), a 50M-example dataset designed for instruction fine-tuning and pre-training of speech-text large language models (LLMs). SIFT-50M is built from publicly available speech corpora, which collectively contain 14K hours of speech, and leverages LLMs along with off-the-shelf expert models. The dataset spans five languages, encompassing a diverse range of speech understanding as well as controllable speech generation instructions. Using SIFT-50M, we train SIFT-LLM, which outperforms existing speech-text LLMs on instruction-following benchmarks while achieving competitive performance on foundational speech tasks. To support further research, we also introduce EvalSIFT, a benchmark dataset specifically designed to evaluate the instruction-following capabilities of speech-text LLMs.

\end{abstract}

\section{Introduction}

Recent years have witnessed significant advancements in integrating speech and audio capabilities into large language models (LLMs). A common approach involves coupling an audio encoder with an LLM by projecting audio embeddings into the model’s input space \cite{tangsalmonn,yang2024qwen2,gong2023joint,das2024speechverse}. This integration preserves rich prosodic and acoustic information beyond textual transcriptions, enabling speech-text LLMs to perform a broad range of speech understanding tasks. Another research direction focuses on extending LLM vocabulary to generate discrete audio tokens for speech synthesis \cite{rubenstein2023audiopalm,du2023lauragpt}.

Speech-text LLMs are typically trained on existing task-specific speech corpora, which are primarily designed for automatic speech recognition (ASR). These datasets predominantly consist of paired audio and task-specific labels rather than natural language instructions, limiting their utility for instruction-following training. This lack of diverse, large-scale instruction datasets poses challenges in generalizing to broader speech understanding tasks. To address this gap, we introduce SIFT-50M\footnote{The dataset can be accessed at \url{https://huggingface.co/datasets/amazon-agi/SIFT-50M}} (\textbf{S}peech \textbf{I}nstruction \textbf{F}ine-\textbf{T}uning), a large-scale multilingual dataset covering five languages for instruction-based speech model training. SIFT-50M augments existing speech datasets with instruction-based question-answer (QA) pairs for speech understanding and includes approximately 5M examples for controllable speech generation.

In summary, our main contributions are as follows: (1) We release SIFT-50M, a 50M-example multilingual instruction dataset which, to the best of our knowledge, is the largest publicly available instruction dataset for speech understanding tasks. (2) We introduce EvalSIFT, a benchmark dataset designed for systematically evaluating speech-text LLMs across multiple dimensions. (3) We develop SIFT-LLM, a speech-text LLM trained on SIFT-50M, achieving state-of-the-art performance on instruction-following benchmarks compared to publicly available models of similar size. (4) We conduct controllable speech generation experiments using SIFT-50M, demonstrating the model’s ability to generate speech in instructed styles.

\section{Related Work}

\textbf{Speech-text LLMs:} Several speech-text LLMs have been introduced recently. Early works focused on specific tasks such as ASR \cite{fathullah2024prompting, yu2024connecting} and second-pass ASR rescoring \cite{li2023prompting}. SALMONN \cite{tangsalmonn} integrates Whisper \cite{radford2023robust} and BEATs \cite{chen2023beats} encoders with a pre-trained LLM using a window-level Q-former \cite{li2023blip}. Qwen-Audio \cite{chu2023qwen} and Qwen2-Audio \cite{chu2024qwen2} are trained on over 30 speech, audio, and music tasks. WavLLM \cite{hu2024wavllm} incorporates two speech encoders, Whisper-large-v2 and WavLM-base \cite{chen2022wavlm}, while LTU-AS \cite{gong2023joint} integrates Whisper-large with Llama-7B \cite{touvron2023llama} via a time and layerwise transformer \cite{gong2023whisper}. These models process continuous audio representations, projecting encoded audio into the LLM’s input space.

To support both understanding and generation, AudioPaLM \cite{rubenstein2023audiopalm} and LauraGPT \cite{du2023lauragpt} extend LLM vocabulary with discrete audio tokens. AudioPaLM discretizes embeddings from USM \cite{zhang2023google} and w2v-BERT \cite{chung2021w2v} using k-means clustering, extending LLM vocabulary with discrete audio tokens. LauraGPT adopts a hybrid approach, utilizing a conformer-based encoder for continuous representation of input speech while generating discrete EnCodec \cite{defossez2022high} codes for output speech.

\textbf{Instruction Fine-Tuning Datasets:} Most prior speech-text LLMs rely on task-specific datasets with instructions generated by LLMs. To improve task diversity, WavLLM \cite{hu2024wavllm} created a 3K-hour speech question-answering dataset using GPT-4 based on transcriptions from Librispeech \cite{panayotov2015librispeech}, the AMI Meeting Corpus \cite{carletta2005ami}, Fisher \cite{cieri2004fisher}, and Switchboard \cite{godfrey1992switchboard}. Additionally, they developed a multi-task dataset by merging independent training examples into single multi-task instructions.

Another relevant dataset is OpenASQA \cite{gong2023joint}, a publicly available instruction fine-tuning dataset with 2.7M examples for speech and audio tasks. Its construction methodology, closely related to ours, leverages datasets such as IEMOCAP \cite{busso2008iemocap}, LibriTTS \cite{zen2019libritts}, VoxCeleb2 \cite{chung2018voxceleb2}, and MOSEI \cite{zadeh2018multimodal}. OpenASQA augments existing annotations (e.g., transcription, emotion, gender) with metadata such as speed, pitch, and energy, and then uses this enriched metadata to prompt GPT-3.5-Turbo to generate QA pairs. In contrast, SIFT-50M is an order of magnitude larger than OpenASQA, covering a broader range of speech dimensions (e.g., accent, age, gender, pitch, speaking rate, word alignment, room characteristics, noise level, distortion) while also supporting instructions in four additional languages beyond English.

\textbf{Controllable Speech Generation}: Prompt-based speech generation, enabling controllable synthesis of acoustic characteristics such as gender, speaking rate, pitch, and loudness, has become a prominent area of investigation \cite{guo2023prompttts, shimizu2024prompttts++}. \cite{kawamura2024libritts} annotated the LibriTTS corpus \cite{zen2019libritts} with textual prompts to augment these specific speech properties. While \cite{lyth2024natural} detailed a method for annotating a substantial 45K-hour corpus for high-fidelity speech generation, these prompts are not publicly available, limiting reproducibility and benchmarking. Conversely, SIFT-50M offers speech generation instructions across five languages, enabling control over a broader spectrum of parameters, including gender, accent, age, speaking rate, pitch, and intensity.

\section{SIFT Dataset Construction}
\begin{figure*}[t]  % Use figure* for spanning both columns
    \centering
    \includegraphics[width=1.0\textwidth, height=0.5\textheight, keepaspectratio]{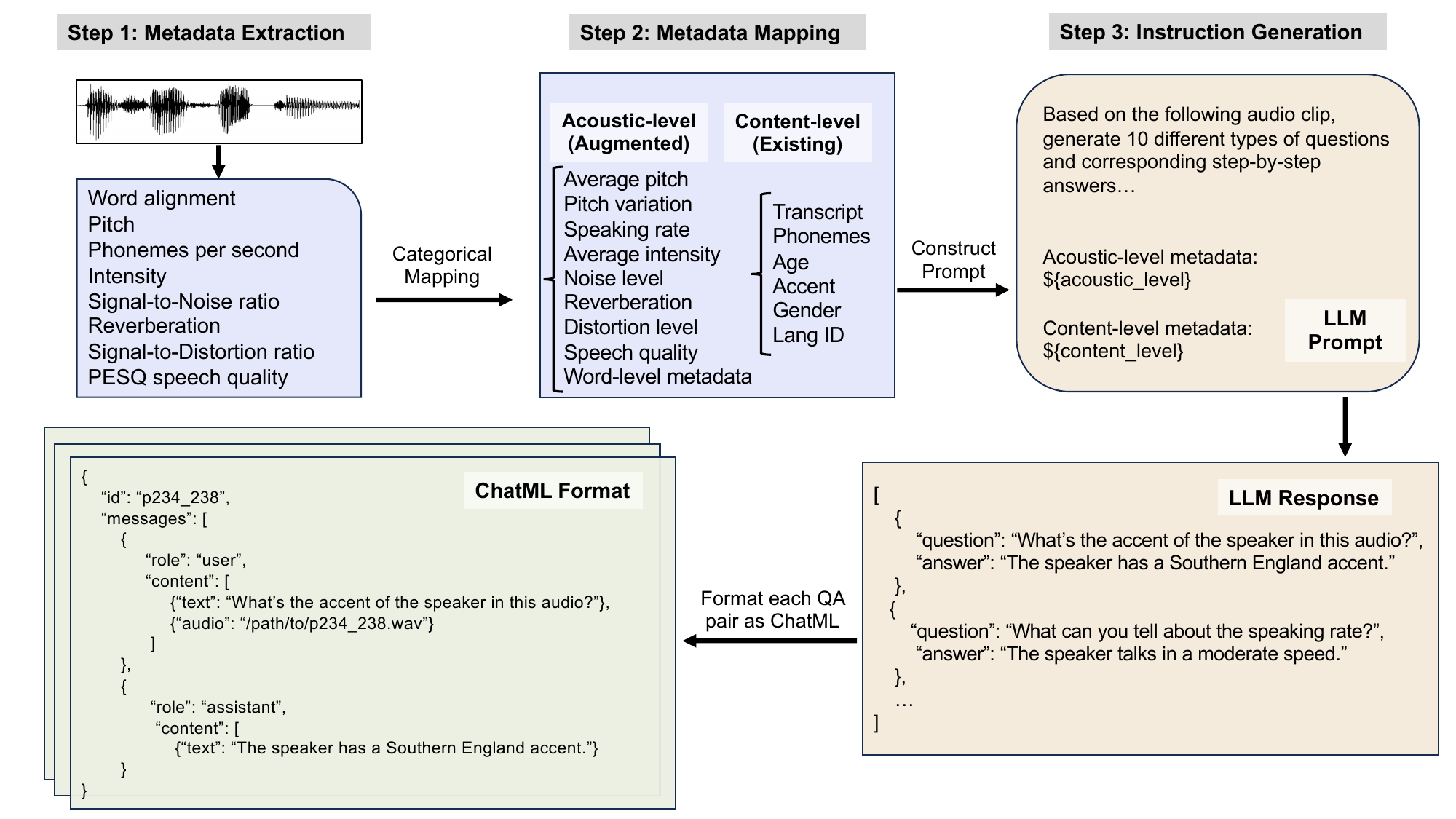}  % Adjust width as needed
    \vspace{-0.4cm}
    \caption{Block diagram showing the stages of SIFT-50M dataset construction. For non-English data generation, we substitute the metadata mapping with the respective language and prompt the LLM to generate responses in that language.}
    \label{fig:block_diag}
% \vspace{-2mm}
\end{figure*}

To construct SIFT-50M, we utilize three publicly available multilingual speech data sources: Multi-lingual Librispeech \cite{pratap2020mls}, Common Voice \cite{ardila2019common}, and VCTK \cite{Yamagishi2019CSTRVC}. These sources provide a diverse range of speakers, speech styles, languages, and accents. We augment the existing metadata with additional features extracted from the speech data and generate natural language instructions from the enriched metadata with the help of an LLM. Finally, we employ a feedback loop involving verification and iterative prompt refinement to improve the generation process.

 % These steps are detailed in the sub-sections below. 

\subsection{Metadata Extraction}
\label{sec:metadata}
We augment the source speech datasets—which typically include content-level characteristics, with additional acoustic-level features such as pitch, speaking rate, intensity, duration, reverberation, and noise level. We extract pitch using PYYAPT \cite{kasi2002yet} and intensity (in decibels) using Parselmouth \cite{jadoul2018introducing}. Phonemes are derived from transcripts via a grapheme-to-phoneme (G2P) model \cite{g2pE2019}, while noise, speech quality, and reverberation features are computed following the method described in \cite{dataspeech}. Content-level attributes (e.g., transcription, gender, age, and accent) are provided in the source datasets, although accent and age information is available only for a random subset of utterances in the Common Voice and VCTK datasets. When a gender label is missing, we employ a gender classification model \cite{voice_gender} to fill in the gap.

We then convert numerical utterance-level statistics into categorical ranges for pitch, intensity, age, reverberation, noise level, and speech quality based on the distribution of each variable within a dataset. For example, pitch is binned separately for male and female speakers due to their different typical ranges. Specifically, for a male speaker, an \textit{average pitch} metadata is mapped to \textit{slightly low pitch} when the median pitch falls within $125 < f_{\text{median}} < 138$ Hz, while for a female speaker the corresponding range is $166 < f_{\text{median}} < 182$ Hz. Age is similarly grouped into broad categories (e.g., twenties, thirties, …, nineties) to both de-identify and generalize the attribute. For a detailed list of categorical values, please refer to Appendix Section \ref{sec:appendix_metadata_map}.

To generate word-level metadata, we extract time alignments for each word using our in-house hybrid HMM-LSTM acoustic model \cite{sak2014long} for forced alignment. The start and end times of each word are then used to derive word-level pitch and intensity statistics, which are subsequently converted to categorical values. For non-English languages (French, German, Spanish, and Italian), we map the metadata characteristics and their corresponding categorical values to the target language using a phrase-level translation dictionary.

\begin{table}[h]
  \centering
  \small
  \renewcommand{\arraystretch}{1.1} % Increase row spacing by 50%
  \begin{tabular}{p{7.0cm}}
    \toprule
    \textbf{Closed-ended: Acoustic-Level} \\
    \textbf{Question}: How would you characterize the speaking rate in this audio?\\
    \textbf{Answer}: The speaking rate in this audio is slightly slow.\\
    \hline 
    \textbf{Closed-ended: Word-Align}       \\
    \textbf{Question}:  What is the average pitch for the word `bent' and how does it compare to the average pitch of the surrounding words?\\
    \textbf{Answer}:   The average pitch for the word `bent' is quite low. The previous word `upon' has a similar average pitch, while the following word `upon' has a slightly lower pitch. Overall, there is not a significant change in pitch for these words.\\
    \hline
    \textbf{Closed-ended: Comparison}        \\
    \textbf{Question}:  Which audio has a speaker with a Scottish accent?\\
    \textbf{Answer}:  The second audio has a speaker with a Scottish accent, whereas the first audio has a speaker with an English accent.\\
    \hline
    \textbf{Open-ended}  \\
    \textbf{Question}:   Analyze the phonetic structure of the audio and discuss how the phonemes contribute to the overall perception of the speaker's accent and age category.\\
    \textbf{Answer}: The phonetic structure, including phonemes like `IH1', `T', `S', `IY1', `M', '`', `AH0', `W', `EY1', `S', `T', `AH1', `V', `T', `AE1', `L', `AH0', `N', `T', `S', reflects an Irish accent characterized by distinct vowel sounds and consonant clusters. The precise articulation and the youthful intonation patterns suggest the speaker is in their twenties.\\
    \hline
    \textbf{Controllable Generation}  \\
    \textbf{Question}: Please synthesize a speech sample with a very low pitch and quite monotone pitch variation, spoken quite slowly and at a moderate volume. The speaker should be an adult woman in her fifties with a United States English accent. Given the transcription text as: In my book of memory, time writes tales of moments past. \\
    \textbf{Answer}: \textit{<reference to original audio>} \\
    \bottomrule
  \end{tabular}
  \caption{Representative examples from SIFT-50M for different categories of instruction types.}
  \label{tab:qa_types}
\vspace{-4mm}
\end{table}

\begin{table}[h]
  \centering
  \small
  \renewcommand{\arraystretch}{1.05} % Increase row spacing by 50%
  \begin{tabular}{p{3.0cm}p{2.8cm}}
    \toprule
    \textbf{Category} & {\begin{minipage}[t]{3cm}\centering \textbf{\#Samples} (train/dev/EvalSIFT)\end{minipage}} \\
    \midrule
    Closed-Ended & \\
    \hspace{0.4cm} -- Acoustic-level   & {\begin{minipage}[t]{3cm}\centering 17.8M / 100K / 2.5K\end{minipage}} \\
    \hspace{0.4cm} -- Content-level    & {\begin{minipage}[t]{3cm}\centering 14.5M / 80K / 2.5K\end{minipage}} \\
    \hspace{0.4cm} -- Word-Align       & {\begin{minipage}[t]{3cm}\centering 9.8M / 40K / 2.5K\end{minipage}} \\
    \hspace{0.4cm} -- Comparison       & {\begin{minipage}[t]{3cm}\centering 3.6M / 100K / 2.5K\end{minipage}} \\
    Open-Ended & {\begin{minipage}[t]{3cm}\centering 4.3M / 100K / 10K\end{minipage}} \\
    Controllable Generation & {\begin{minipage}[t]{3cm}\centering 5.6M / 50K / 10K\end{minipage}} \\
    \midrule
    \textbf{Total} & {\begin{minipage}[t]{3cm}\centering \textbf{55.6M / 470K / 30K}\end{minipage}} \\
    \bottomrule
  \end{tabular}
  \caption{High-level statistics of SIFT-50M dataset showing the distribution of categories pooled across the source speech datasets and languages.}
  \label{tab:data_volume}
\vspace{-4mm}
\end{table}

\subsection{Natural Language Instruction Generation}
Our instructions are organized into three main categories: closed-ended, open-ended, and controllable generation. To generate instruction-response pairs, we feed categorical metadata to the LLM and prompt it to produce up to 10 QA pairs per speech utterance, as shown in Figure~\ref{fig:block_diag}. For multilingual instruction generation, the LLM is prompted to generate instructions in the specified language using the corresponding mapped metadata. Table~\ref{tab:qa_types} provides examples of QA pairs for different categories in the SIFT-50M dataset. High-level statistics of SIFT-50M are presented in Table~\ref{tab:data_volume}.

\textbf{Closed-ended Instructions:}  
These instructions are subdivided into four sub-categories: (1) Acoustic-Level, (2) Content-Level, (3) Word-Align, and (4) Comparison types. The first three sub-categories, generated using Mixtral 8x7B \cite{jiang2024mixtral}, are based on the respective metadata types. As reported in LTU \cite{gong2023listen}, closed-ended instructions are crucial for guiding the model to understand speech and follow instructions without hallucinating. Unlike LTU’s OpenASQA data, which relies on pre-processed questions and strict answer templates, our approach leverages careful prompt engineering to allow the model to construct diverse and generalizable QA pairs. In addition, a language field is incorporated into the metadata (by sampling examples from the top-20 locales in the Common Voice dataset) to facilitate the generation of instructions for the acoustic-based language identification (LID) task.

We introduce a novel task in the form of \textit{Comparison} instructions, where the LLM compares two audio files based on their speech characteristics. The LLM is provided with metadata from two distinct audio files and is prompted to generate comparative questions and free-form answers rather than binary responses. Due to the increased complexity of this task, we employ Amazon Nova Pro \cite{AmazonAGI2024}, which has demonstrated stronger performance on text benchmarks compared to Mixtral 8x7B.\\
\textbf{Open-ended Instructions:}  
In this category, the LLM is prompted to generate more complex and diverse questions, with answers inferred from the metadata rather than provided directly by it. This approach encourages the model to produce thoughtful and detailed responses. We employ Amazon Nova Pro to generate open-ended dataset.\\
\textbf{Controllable Generation:}
This category includes  instructions for expressive speech synthesis from text with specific speech characteristics. These instructions are generated using Mixtral 8x7B based on the provided acoustic and content-level metadata.\\
\textbf{EvalSIFT:} We also release a benchmark, EvalSIFT, constructed in a manner similar to SIFT-50M. For each of the five languages, we generate 2K examples per category (closed-ended, open-ended, and controllable generation), yielding a total of 30K examples.

\subsection{Quality Assurance}
During the metadata extraction stage, we reject any values that fall outside the feasible ranges for the respective speech characteristics, ensuring accurate mapping from numerical to categorical values. High-quality, well-crafted LLM-generated data is essential for efficient model fine-tuning and improving the model's ability to follow unseen instructions. We employ two types of quality control over the LLM-generated datasets. First, the authors perform an exhaustive review of the LLM responses across all categories and languages. We establish a feedback loop with iterative prompt refinement to ensure that the model's responses are cogent, meet our expectations, and are free from hallucinations. Second, we conduct ablation studies on the generated data by evaluating performance on the SIFT-50M development sets, which allows us to identify and discard problematic data or even entire categories. For example, we found that instructions based on speech-text alignment that include numerical values (e.g., “What words occur between the 3 and 5 second mark?”) significantly degraded model performance. Consequently, we retain only instructions with non-numerical values (i.e., instructions similar to the word-align category shown in Table~\ref{tab:qa_types}). We release the rejected portion as a bonus research partition to support future research directions.

\section{Experimental Setup}
To evaluate the effectiveness of SIFT-50M, we train a speech-text LLM, called SIFT-LLM, on our dataset. In this section, we detail the model architecture and training configuration. 

\subsection{SIFT-LLM Architecture}\label{sec:sift_llm}

\textbf{Acoustic Encoders:} We adopt a hybrid speech representation approach: input speech is encoded as continuous embeddings, while output speech is generated as discrete tokens. Specifically, we use Whisper-medium \cite{radford2023robust} as the encoder to process input audio. For discrete speech tokens, we experiment with two codec options: (1) k-means clustered HuBERT embeddings \cite{lakhotia2021generative}, which primarily capture the semantic content of speech, and (2) X-codec2 \cite{ye2025llasa}, which fuses semantic and acoustic codes into a unified codebook. The HuBERT codes are subsequently converted to audio using a vocoder based on the Big-VGan architecture \cite{lee2023bigvgan}. Unless stated otherwise, the setup should be assumed to use HuBERT codes.\\
\textbf{Large Language Model:} Our core LLM is Qwen2.5-7B-instruct \cite{yang2024qwen2}. We add a linear layer on top of the Whisper-medium encoder to project the $1024$-dimensional speech embeddings into the LLM’s $3584$-dimensional space. Additionally, we expand the LLM's vocabulary by the size of the codebook ($2000$ for HuBERT codes and $65536$ for X-codec2). We employ low-rank adaptation \cite{hu2021lora} to train the LLM parameters.

\subsection{Model Training}
\label{sec:model_training}
\textbf{Continued Pre-training:} To align the audio embedding space of the pre-trained Whisper encoder with the input representation of the Qwen2.5 LLM and to train the randomly initialized projection layer, we perform continued pre-training of SIFT-LLM. This phase involves training on a mixture of speech understanding and generation tasks—namely, ASR, emotion recognition (ER), speaker-attributed ASR (SA-ASR), intent classification (IC), slot entity recognition (SER), speech-to-text translation (S2TT), speech-to-speech translation (S2ST), and text-to-speech (TTS). We convert the IC task into a multiple-choice question format by providing possible intent values in the instructions. All tasks are represented using natural language instructions. In addition to SIFT-50M, we also release the prompt templates used during this stage.

During pre-training, the acoustic encoder remains frozen, and we train only the linear connector layer and the LoRA parameters of the LLM. We set the LoRA rank to 16, resulting in $8.7$M learnable parameters, and train for a total of 200K steps. The resulting pre-trained models are referred to as SIFT-LLM PT. For controllable generation experiments, we continue pre-training for an additional 200K steps, using a LoRA rank of 128 and making the core LLM's embedding layer fully trainable. In this stage, the total number of learnable parameters increases to $590$M for HuBERT codes setup and $830$M for X-codec2 setup. We also increase the relative weights of the TTS and S2ST datasets.

\textbf{Instruction Fine-tuning}: Following continued pre-training, we perform an instruction fine-tuning stage using the SIFT-50M dataset. We conduct separate fine-tuning processes for speech understanding and controllable generation. For speech understanding, training is initialized from the 50K\textsuperscript{th} SIFT-LLM PT checkpoint and uses only the speech understanding instructions (closed- and open-ended) from SIFT-50M. The number of trainable parameters remains the same as during pre-training for the first 200K steps. We refer to this model as SIFT-LLM. For controllable generation, training is initialized from the 400K\textsuperscript{th} SIFT-LLM PT checkpoint and focuses exclusively on controllable generation instructions, with the number of trainable parameters unchanged from the last 200K steps of pre-training. This model is referred to as SIFT-LLM GEN. For further details on training datasets, prompts, and setup, we refer the reader to Appendix~\ref{appendix:training_setup}.

\begin{table*}[t]
    \centering
    \small
    \renewcommand{\arraystretch}{1.1}
    \begin{tabular}{p{2.5cm}|p{0.8cm}p{1.2cm}|p{1.1cm}p{1.1cm}}
        \toprule
        \multirow{2}{*}{\textbf{Model}} &  \multicolumn{2}{c}{\textbf{Closed-Ended}} & \multicolumn{2}{|c}{\textbf{Open-Ended}}\\
&
\begin{minipage}[t]{0.8cm}\centering DS-1 \end{minipage} &
\begin{minipage}[t]{1.2cm}\centering EvalSIFT\end{minipage} &
\begin{minipage}[t]{1.2cm}\centering AB-Chat\end{minipage} &
\begin{minipage}[t]{1.1cm}\centering EvalSIFT\end{minipage} \\
        \midrule
SALMONN-7B &
\begin{minipage}[t]{0.8cm}\centering 34.7 \end{minipage} &
\begin{minipage}[t]{1.2cm}\centering 21.9 \end{minipage} &
\begin{minipage}[t]{1.2cm}\centering 6.4 \end{minipage} &
\begin{minipage}[t]{1.1cm}\centering 6.0 \end{minipage} \\
Qwen2-Audio-Inst. &
\begin{minipage}[t]{0.8cm}\centering \underline{48.0} \end{minipage} &
\begin{minipage}[t]{1.2cm}\centering \underline{25.1} \end{minipage} &
\begin{minipage}[t]{1.2cm}\centering \underline{7.2} \end{minipage} &
\begin{minipage}[t]{1.1cm}\centering \underline{7.3} \end{minipage} \\
O-ASQA-LLM &
\begin{minipage}[t]{0.8cm}\centering 45.9 \end{minipage} &
\begin{minipage}[t]{1.2cm}\centering 22.9 \end{minipage} &
\begin{minipage}[t]{1.2cm}\centering 6.6 \end{minipage} &
\begin{minipage}[t]{1.1cm}\centering 4.7  \end{minipage} \\
SIFT-LLM (ours) &
\begin{minipage}[t]{0.8cm}\centering \textbf{57.4} \end{minipage} &
\begin{minipage}[t]{1.2cm}\centering \textbf{46.1} \end{minipage} &
\begin{minipage}[t]{1.2cm}\centering \textbf{7.3} \end{minipage} &
\begin{minipage}[t]{1.1cm}\centering \textbf{7.8} \end{minipage} \\
\bottomrule
    \end{tabular}
    \hspace{0.15cm}
    \begin{tabular}{p{0.6cm}p{0.6cm}p{0.6cm}p{0.7cm}p{0.7cm}p{0.8cm}}
        \toprule
        \multicolumn{6}{c}{\textbf{Dynamic-Superb Tasks}} \\
\begin{minipage}[t]{0.6cm}\centering Audio \end{minipage} & 
\begin{minipage}[t]{0.6cm}\centering PL \end{minipage} & 
\begin{minipage}[t]{0.6cm}\centering Semt. \end{minipage} & 
\begin{minipage}[t]{0.7cm}\centering Degrd. \end{minipage} & 
\begin{minipage}[t]{0.7cm}\centering Content \end{minipage} & 
\begin{minipage}[t]{0.8cm}\centering Speaker \end{minipage} \\
        \midrule
\begin{minipage}[t]{0.6cm}\centering 31.7 \end{minipage} & 
\begin{minipage}[t]{0.6cm}\centering \underline{30.5} \end{minipage} & 
\begin{minipage}[t]{0.6cm}\centering \underline{47.5} \end{minipage} & 
\begin{minipage}[t]{0.7cm}\centering 30.0 \end{minipage} & 
\begin{minipage}[t]{0.7cm}\centering 45.2 \end{minipage} & 
\begin{minipage}[t]{0.8cm}\centering 31.9 \end{minipage}
\\
\begin{minipage}[t]{0.6cm}\centering \textbf{53.5} \end{minipage} & 
\begin{minipage}[t]{0.6cm}\centering 28.9 \end{minipage} & 
\begin{minipage}[t]{0.6cm}\centering 40.3 \end{minipage} & 
\begin{minipage}[t]{0.7cm}\centering 43.9 \end{minipage} & 
\begin{minipage}[t]{0.7cm}\centering 70.6 \end{minipage} & 
\begin{minipage}[t]{0.8cm}\centering \underline{43.6} \end{minipage}
\\
\begin{minipage}[t]{0.6cm}\centering 28.5 \end{minipage} & 
\begin{minipage}[t]{0.6cm}\centering 30.0 \end{minipage} & 
\begin{minipage}[t]{0.6cm}\centering 38.6 \end{minipage} & 
\begin{minipage}[t]{0.7cm}\centering \underline{45.9} \end{minipage} & 
\begin{minipage}[t]{0.7cm}\centering \underline{72.3} \end{minipage} & 
\begin{minipage}[t]{0.8cm}\centering 40.7 \end{minipage}
\\
\begin{minipage}[t]{0.6cm}\centering \underline{37.5} \end{minipage} & 
\begin{minipage}[t]{0.6cm}\centering \textbf{42.8} \end{minipage} & 
\begin{minipage}[t]{0.6cm}\centering \textbf{51.3} \end{minipage} & 
\begin{minipage}[t]{0.7cm}\centering \textbf{63.6} \end{minipage} & 
\begin{minipage}[t]{0.7cm}\centering \textbf{75.6} \end{minipage} & 
\begin{minipage}[t]{0.8cm}\centering \textbf{47.7} \end{minipage}\\
\bottomrule
    \end{tabular}
\caption{Evaluation results of speech-text LLMs on Dynamic-Superb (DS-1), AIR-Bench Chat (AB-Chat), and EvalSIFT (English). We report accuracy (in \%) for closed-ended evaluations and LLM score (0 to 10) for open-ended evaluations. The adjacent table provides a breakdown by task categories in Dynamic-Superb -- Audio, Paralinguistics (PL), Semantics (Semt.), Degradation (Degrd.), Content, and Speaker. \textbf{Bold} values indicate the best results, and \underline{underlined} values indicate the second-best results.}
\label{tab:if_results}
% \vspace{-1mm}
\end{table*}

\begin{table*}[t]
  \centering
  \small
  \renewcommand{\arraystretch}{1.1} % Increase row spacing by 50%
  \begin{tabular}{p{1.7cm}|c|cccc|cc}
    \toprule
\multirow{2}{1.7cm}{\begin{minipage}[t]{1.7cm}\centering \textbf{Task (Metric)} \end{minipage}} &
\multirow{2}{*}{\textbf{Test Set}} &
\multirow{2}{*}{ \begin{minipage}[t]{1.4cm}\centering \textbf{SALMN} \end{minipage} } &
\multirow{2}{*}{ \begin{minipage}[t]{1.4cm}\centering \textbf{QwA-Inst} \end{minipage} } &
\multirow{2}{*}{ \begin{minipage}[t]{1.4cm}\centering \textbf{OASQAL} \end{minipage} } &
\multirow{2}{*}{\textbf{SIFT-LLM}} &
\multicolumn{2}{c}{\textbf{SIFT-LLM PT}} \\

    & & & & & & 50k-ckp & 200k-ckp \\
    \midrule
\multirow{5}{1cm}{\begin{minipage}[t]{1.25cm}\centering ASR (WER $\downarrow$)\end{minipage}} & LS-Clean &
\textbf{2.5}  &
4.8 &
3.8 &
\underline{3.5} & 2.5 & 2.3\\

& LS-Other &
\textbf{5.7}  &
\begin{minipage}[t]{1.5cm}\centering \underline{7.4} \end{minipage} &
\begin{minipage}[t]{1.5cm}\centering 8.1 \end{minipage} &
7.5 & 5.4 & 5.0\\
& PS  &
\underline{22.2} &
\begin{minipage}[t]{1.5cm}\centering 29.4 \end{minipage} &
\begin{minipage}[t]{1.5cm}\centering \textbf{19.4} \end{minipage} &
26.0 & 24.4 & 24.9 \\

& FLEURS-en  &
9.2  &
\begin{minipage}[t]{1.5cm}\centering \underline{9.1} \end{minipage} &
\begin{minipage}[t]{1.5cm}\centering 11.3 \end{minipage} &
\textbf{8.1} & 6.6 & 6.4 \\

& FLEURS-5  &
23.7  &
\begin{minipage}[t]{1.5cm}\centering \underline{13.2} \end{minipage} &
\begin{minipage}[t]{1.5cm}\centering 23.1 \end{minipage} &
\textbf{11.4} & 9.0 & 8.2 \\

\midrule
\multirow{2}{1cm}{\begin{minipage}[t]{1.5cm}\centering ER (Acc. $\uparrow$) \end{minipage}} &
MSP-test1    & 38.5  & \underline{40.0} & 38.6 & \textbf{53.6} & 54.3 & 54.3\\
& MSP-test2  & 28.4  & 32.1 & \underline{40.4} & \textbf{50.2} & 52.4 & 52.7\\

\midrule
\begin{minipage}[t]{1.5cm}\centering IC (Acc. $\uparrow$)\end{minipage} & SLURP  &
58.4  &
\begin{minipage}[t]{1.5cm}\centering \underline{86.0} \end{minipage} &
\begin{minipage}[t]{1.5cm}\centering 68.0 \end{minipage} &
\textbf{92.7} & 94.8 & 96.0 \\
\begin{minipage}[t]{1.7cm} SER (Acc. $\uparrow$)\end{minipage} & SLURP  &
26.4 &
\begin{minipage}[t]{1.5cm}\centering \underline{52.1} \end{minipage} &
\begin{minipage}[t]{1.5cm}\centering 48.3 \end{minipage} &
\textbf{71.3} & 73.1 & 72.7 \\

\midrule
\multirow{4}{1cm}{\begin{minipage}[t]{1.5cm}\centering S2TT (BLEU $\uparrow$)\end{minipage}} & de $\rightarrow$ en &
21.3 &
\begin{minipage}[t]{1.5cm}\centering \textbf{31.4} \end{minipage} &
\begin{minipage}[t]{1.5cm}\centering 28.9 \end{minipage} &
\underline{29.2} & 34.4 & 35.2\\

& fr $\rightarrow$ en &
20.6 &
\begin{minipage}[t]{1.5cm}\centering \textbf{34.2} \end{minipage} &
\begin{minipage}[t]{1.5cm}\centering 29.8 \end{minipage} &
\underline{30.9} & 36.3 & 37.0 \\

& it $\rightarrow$ en &
18.9 &
\begin{minipage}[t]{1.5cm}\centering \textbf{33.6} \end{minipage} &
\begin{minipage}[t]{1.5cm}\centering 29.4 \end{minipage} &
\underline{31.7} & 36.0 & 36.8\\

& es $\rightarrow$ en &
21.6 &
\begin{minipage}[t]{1.5cm}\centering \textbf{36.3} \end{minipage} &
\begin{minipage}[t]{1.5cm}\centering 33.1 \end{minipage} &
\underline{35.5} & 39.7 & 40.4\\
\bottomrule
  \end{tabular}
\caption{Evaluation results for SALMONN-7B (SALMN), Qwen2-Audio-Instruct (QwA-Inst), O-ASQA-LLM (OASQAL), and SIFT-LLM on foundational tasks, along with results for SIFT-LLM PT $50K$ and $200K$ checkpoints. Accuracies (Acc.) are reported as percentages. FLEURS-5 includes test sets for five languages: en, de, fr, it, and es. PS denotes the People's Speech dataset. S2TT results are reported on CoVoST2. \textbf{Bold} values indicate the best results, while \underline{underlined} values represent the second-best results among \textbf{instruction fine-tuned models}.}
  \label{tab:core_results}
\vspace{-2mm}
\end{table*}

\section{Experimental Results}
\label{sec:experiments}

\subsection{Evaluation Setup}
We evaluate SIFT-LLM by benchmarking it against publicly available speech-text LLMs of similar sizes, including SALMONN-7B and Qwen2-Audio-7B-Instruct. For a direct comparison between the SIFT-50M dataset and the OpenASQA dataset, we train a speech-text LLM using the same pre-trained model as SIFT-LLM but fine-tuned on OpenASQA; we refer to this model as O-ASQA-LLM. Across all evaluations, we use Temperature$=0.1$, Top\ P$=0.95$, a repetition penalty of $1.1$ \cite{keskar2019ctrl}, and set the no\_repeat\_ngram size to 3 \cite{wolf2020transformers}.

\subsection{Instruction-following Evaluation}
\textbf{Benchmarks:} Several benchmarks have been proposed for evaluating instruction-following speech-text LLMs. For example, Dynamic-Superb \cite{huang2024dynamic} originally comprised 55 speech and audio tasks and was later expanded to 180 tasks in its Phase-2 release \cite{huang2024dynamic2}. AIR-Bench \cite{yang2024air} is composed of two parts: \textit{Foundation}, which covers 19 standard speech tasks, and \textit{Chat}, which includes open-ended questions. We use Dynamic-Superb (DS-1)\footnote{Phase-2 data was not available at the time of writing.} for closed-ended evaluations, and the \textit{Chat} partition of AIR-Bench for open-ended evaluation, selecting only the speech subset (excluding sound and music). Additionally, we report metrics on both the closed- and open-ended categories of EvalSIFT.

SIFT-LLM and O-ASQA-LLM accept any number of text and audio segments in all possible permutations, whereas Qwen2-Audio-Instruct and SALMONN-7B accept only a single audio input. To address this limitation during evaluation for comparison tasks in EvalSIFT or for tasks in Dynamic-Superb that involve multiple audio segments, we concatenate the segments with a 1-second silence between each.

\textbf{Metrics:} While exact match accuracy is a common metric for closed-ended or classification tasks, it becomes less reliable for speech-text LLMs, which tend to produce verbose, open-ended responses. Therefore, we adopt an LLM-as-a-judge strategy \cite{zheng2023judging} similar to that used in DS-1 and AIR-Bench, employing Claude-3.5-Sonnet\footnote{\url{https://aws.amazon.com/bedrock/claude/}} for evaluation. Using a chain-of-thought reasoning prompt \cite{wei2022chain}, we provide Claude-3.5-Sonnet with both the question and the reference answer. For closed-ended and classification benchmarks, we prompt it to classify responses as correct or incorrect, while for open-ended benchmarks, we ask it to score the responses on a scale from 0 to 10 based on their alignment with the reference answer. We repeated LLM scoring multiple times and found that the width of 95\% confidence interval for the mean accuracy (in \%) was less than 0.1 and for the mean score was less than 0.01.

\textbf{Results:} Table~\ref{tab:if_results} summarizes the performance of various models on instruction-following benchmarks. SIFT-LLM achieves competitive results across all benchmarks and outperforms all other models. Notably, it performs better than O-ASQA-LLM on every benchmark, underscoring the importance of the large-scale and diverse SIFT-50M dataset. Furthermore, SIFT-LLM is second-best on the Audio task in the Dynamic-Superb benchmark, despite not being explicitly trained on audio tasks. We observed that the performance gap between SIFT-LLM and other models is less for Content-level instructions in EvalSIFT, but significantly larger for Acoustic-level, Word-Align, and Comparison type instructions, highlighting the value of data addressing these dimensions in the SIFT-50M dataset. Finally, the model shows signs of reasoning abilities, as evidenced by the examples in Appendix~\ref{sec:reasoning}, where SIFT-LLM reasons through open-ended questions before arriving at its conclusions.\\
\textbf{Results on non-English languages:} Since both Dynamic-Superb and AIR-Bench Chat benchmarks provide instructions exclusively in English, we use EvalSIFT for multilingual evaluation. The results are reported in Table~\ref{tab:ml_results}. SIFT-LLM outperforms SALMONN-7B and Qwen2-Audio-Instruct on closed-ended evaluations across languages. However, Qwen2-Audio-Instruct remains competitive on open-ended evaluations, as observed with English data as well. Overall, absolute accuracy is lower on non-English languages compared to English. Detailed evaluation results are provided in Appendix~\ref{appendix:detailed_results}.

\begin{table}[t]
  \centering
  \small
  \renewcommand{\arraystretch}{1.1} % Increase row spacing by 50%
  \begin{tabular}{lcccc}
    \toprule
\textbf{Model} & \textbf{German} & \textbf{French} & \textbf{Italian} & \textbf{Spanish}\\
\midrule
SAL & 15.0 | 4.3 & 16.3 | 5.0 & 14.3 | 5.0 & 16.7 | 5.4 \\
QwA  & 18.6 | 6.0 & 18.8 | 6.8 & 18.2 | 7.2 & 21.2 | \textbf{7.3} \\
SIFL  & \textbf{39.0} | \textbf{6.6} & \textbf{34.3} | \textbf{7.1} & \textbf{33.2} | \textbf{7.5} & \textbf{35.6} | 7.0 \\
\bottomrule
  \end{tabular}
 \caption{Evaluation results for SALMONN-7B (SAL), Qwen2-Audio-Instruct (QwA), and SIFT-LLM (SIFL) on EvalSIFT for non-English languages. Accuracy (in \%) for the closed-ended category and LLM score (0–10) for the open-ended category are separated by a pipe (|). \textbf{Bold} values indicate the best results.}

  \label{tab:ml_results}
\vspace{-2mm}
\end{table}

\subsection{Evaluation on Foundational Tasks}
In addition to the instruction-following evaluation, we assess performance on standard speech understanding tasks using task-specific metrics: Word Error Rate (WER) for ASR, accuracy for ER, IC, and SER tasks, and BLEU for S2TT. Since model performance can be sensitive to the prompts used \cite{wang2024audiobench}, we use the prompts recommended by the corresponding model wherever possible. For ER, IC, and SER tasks, we leverage Claude-3.5-Sonnet to classify responses as correct or incorrect, ensuring robustness against model-specific output formats. Sequence generation tasks, such as ASR and S2TT pose additional challenges due to some models producing extra text; to address this, we use regular expressions and Claude-3.5-Sonnet to parse model hypotheses. Furthermore, we apply Whisper text normalization\footnote{\url{https://github.com/openai/whisper/tree/main/whisper/normalizers}} before computing WER.

The results, presented in Table~\ref{tab:core_results}, reveal that no single model excels across all tasks. While SALMONN-7B achieves the best performance on ASR benchmarks for English, Qwen2-Audio-Instruct outperforms the others on S2TT, and SIFT-LLM significantly outperforms on ER and spoken language understanding tasks. On the multilingual ASR task, SIFT-LLM and Qwen2-Audio-Instruct show significantly better results than SALMONN-7B and O-ASQA-LLM. Compared to the pre-trained model from which SIFT-LLM is initialized (the $50K$\textsuperscript{th} checkpoint), SIFT-LLM shows a slight decline in performance across tasks—an observation consistent with findings for Qwen2-Audio and its instruction fine-tuned variant \cite{wang2024audiobench}. Furthermore, performance differences between the $50K$ and $200K$ pre-trained checkpoints are minimal across all tasks.

\begin{table*}[h]
  \centering
  \small
  \renewcommand{\arraystretch}{1.05} % Increase row spacing by 50%
  \begin{tabular}{l@{\hspace{0.8cm}}cc@{\hspace{0.8cm}}cc}
    \toprule
        \multirow{2}{*}{\textbf{Feature}} &  \multicolumn{2}{c}{\textbf{HuBERT codes}} & \multicolumn{2}{c}{\textbf{X-codec2}}\\
    &  MAE ($\downarrow$) & QWK ($\uparrow$) & MAE ($\downarrow$) & QWK ($\uparrow$)\\
    \midrule
    Pitch variation &
    \(0.99 \pm 1.05e{-2}\) &
    \(0.15 \pm 2.17e{-2}\) &
    \(0.50 \pm 0.83e{-2}\) &
    \(0.69 \pm 1.50e{-2}\) \\
   Speaking rate &
   \(0.65 \pm 0.35e{-2}\) &
   \(0.46 \pm 0.49e{-2}\) &
   \(0.24 \pm 1.16e{-2}\) &
   \(0.73 \pm 1.67e{-2}\) \\
   
   Intensity &
   \(0.18 \pm 0.16e{-2}\) &
   \(0.02 \pm 0.57e{-2}\) &
   \(0.11 \pm 0.68e{-2}\) &
   \(0.67 \pm 1.78e{-2}\) \\
   \bottomrule
  \end{tabular}
\caption{Evaluation of SIFT-LLM GEN on the controllable generation subset of EvalSIFT. We report MAE and QWK metrics by comparing the speech characteristics of audio generated by SIFT-LLM GEN against the attributes specified in the instructions. We run 10 inference iterations of SIFT-LLM GEN with the temperature set to $0.8$. For HuBERT codes, we synthesize audio using 10 different speaker combinations per iteration. Results are reported as the mean $\pm$ standard deviation.}
\label{tab:controllable_eval}
\vspace{-3mm}
\end{table*}

\subsection{Evaluation on Controllable Generation}
We provide baseline results on the controllable generation set of EvalSIFT for the SIFT-LLM GEN model. To assess controllability, we compare the acoustic metadata extracted from the generated speech with the controllability parameter specified in the instruction. We report two metrics: Mean Absolute Error (MAE) and Quadratic Weighted Kappa (QWK). The distance-based metrics are computed on values categorized into ordinal groups—for example, speaking rate is scored on a scale ranging from “very slowly” (0) to “very fast” (6). We set the sampling temperature to $0.8$ during the inference to balance diversity and coherence in generation.

The results are presented in Table~\ref{tab:controllable_eval}. We focus on stationary characteristics such as pitch variation, speaking rate, and intensity to isolate the controllability evaluation from the effect of the reference speaker signal used during waveform synthesis in the HuBERT-based setup. The results show that the MAE for each category is less than 1 across both setups, indicating that, on average, predictions deviate from the true category by less than one step. Notably, SIFT-LLM GEN with X-codec2 setup demonstrates significantly better controllability. This is attributed to X-codec2’s ability to encode richer speech representations that captures aspects beyond semantics, including prosody and timbre. It also enables more effective control over speaker-dependent characteristics; for example, we found accuracy for gender controllability to be 95.8\% with the X-codec2 setup.

Further analysis of the limitations of semantic-only representations like HuBERT and the effect of speaker references as input to the vocoder, is provided in Appendix~\ref{appendix:controllable_results}.

\subsection{Ablation Studies}

\begin{table}
  \centering
  \small
  \renewcommand{\arraystretch}{1.1} % Increase row spacing by 50%
  \begin{tabular}{p{2.8cm}cp{1.2cm}p{1.2cm}}
    \toprule
    \textbf{Setup} & \textbf{DS-1} & \begin{minipage}[t]{1.5cm}\centering \textbf{EvalSIFT (Closed)} \end{minipage} & \begin{minipage}[t]{1.5cm}\centering \textbf{EvalSIFT (Open)} \end{minipage}\\
    \midrule
Default    &
57.3 &
\begin{minipage}[t]{1.2cm}\centering 45.4 \end{minipage}
& \begin{minipage}[t]{1.2cm}\centering 8.0 \end{minipage} \\
\hline
Init. from 200K  ckpt  & 52.5  &
\begin{minipage}[t]{1.2cm}\centering 42.3 \end{minipage} &
\begin{minipage}[t]{1.2cm}\centering 7.7 \end{minipage}  \\
No Pre-training  & 56.7  &
\begin{minipage}[t]{1.2cm}\centering 43.7 \end{minipage} &
\begin{minipage}[t]{1.2cm}\centering 7.9 \end{minipage}  \\
\hline
LoRA rank=16  & 58.4  &
\begin{minipage}[t]{1.2cm}\centering 45.1 \end{minipage} &
\begin{minipage}[t]{1.2cm}\centering 8.0 \end{minipage}  \\
LoRA rank=32  & 55.7  &
\begin{minipage}[t]{1.2cm}\centering 43.9 \end{minipage} &
\begin{minipage}[t]{1.2cm}\centering 7.8 \end{minipage}  \\
\hline
No open-ended data  & 54.3  &
\begin{minipage}[t]{1.2cm}\centering 42.2 \end{minipage} &
\begin{minipage}[t]{1.2cm}\centering 6.1 \end{minipage}  \\
No word-align data  & 57.1  &
\begin{minipage}[t]{1.2cm}\centering 41.5 \end{minipage} &
\begin{minipage}[t]{1.2cm}\centering 8.0 \end{minipage}  \\
No comparison data  & 56.1  &
\begin{minipage}[t]{1.2cm}\centering 34.3 \end{minipage} &
\begin{minipage}[t]{1.2cm}\centering 7.3 \end{minipage}  \\
\bottomrule
  \end{tabular}
  \caption{Evaluation results on DS-1 and EvalSIFT (English) for various training setups. The \textit{Default} setup is initialized from the 50K\textsuperscript{th} PT checkpoint and instruction-finetuned using LoRA (rank = 8). Metrics reported are accuracy (in \%) for DS-1 and closed-ended EvalSIFT, and LLM score (0–10) for the open-ended EvalSIFT.}
  \label{tab:if_ablation}
\vspace{-3mm}
\end{table}

\begin{figure}[t]
%\vspace{-2mm}
    \centering
    \includegraphics[width=0.45\textwidth]{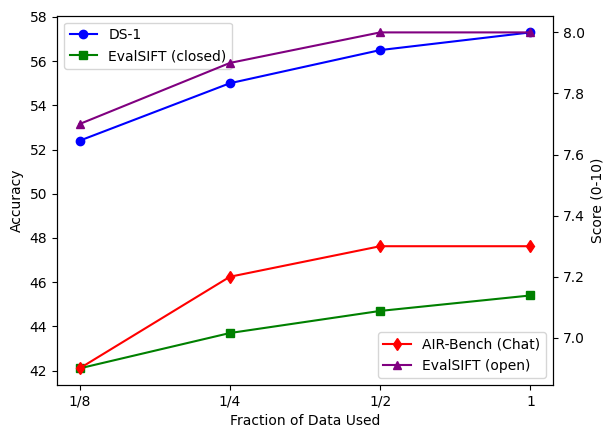}  % Adjust width as needed
    \caption{Effect of SIFT data volume used during instruction fine-tuning on SIFT-LLM's performance, as measured on DS-1, AIR-Bench Chat, and EvalSIFT.}
    \label{fig:data_volume}
\vspace{-3mm}
\end{figure}

For ablation studies, the default SIFT-LLM configuration employs a LoRA rank of 8 and is trained exclusively on the English subset of SIFT-50M during instruction fine-tuning. The remainder of the setup is identical to that described in Section~\ref{sec:model_training}.\\
\textbf{Factors Influencing Instruction-Following:} Table~\ref{tab:if_ablation} summarizes the key factors impacting the model's instruction-following abilities. Extended pre-training can cause the core LLM to lose some of its instruction-following capability. While eliminating pre-training leads to only a moderate degradation on instruction-following benchmarks, it results in a significant decline in performance on foundational tasks like ASR and S2TT. Increasing the LoRA rank from 8 to 16 yields mixed results, but raising it further to 32 results in degradation. Training exclusively on closed-ended datasets significantly impairs open-ended performance, and it also negatively affects accuracy on the closed-ended set. Additionally, omitting comparison data leads to a marked drop in performance on closed-ended EvalSIFT, especially for comparison instructions, since comparison data is the only subset of EvalSIFT that involves multiple audio inputs.\\
\textbf{Effect of Data Volume:} To assess the impact of data volume during instruction fine-tuning, we trained SIFT-LLM using varying fractions of the SIFT-50M dataset: half, one-quarter, and one-eighth of the total volume. The evaluation results, shown in Figure~\ref{fig:data_volume}, indicate that even with just one-eighth of the data, the model achieves strong performance on open-ended benchmarks. However, increasing the data volume leads to further improvements on closed-ended instructions.

\section{Conclusions}
In this work, we introduce SIFT-50M, a large-scale multilingual dataset designed for instruction fine-tuning of speech-text LLMs. By integrating a diverse set of instructions spanning various speech understanding tasks, SIFT-50M not only facilitates instruction-following but also helps models generalize better to unseen instructions. Our experimental results show that SIFT-LLM, our model trained on SIFT-50M, achieves strong performance on instruction-following benchmarks and competitive results on foundational speech understanding tasks. Moreover, we present EvalSIFT, a benchmark dataset tailored for the systematic evaluation of speech-text LLMs for both speech understanding and controllable generation.

\section*{Limitations}
While SIFT-LLM demonstrates strong performance on instruction-following benchmarks, it does not achieve state-of-the-art results on foundational speech tasks. Future work could explore the trade-off between enhancing instruction-following generalization and maintaining competitive performance on foundational tasks relative to task-specific models. Additionally, SIFT-LLM occasionally generates hallucinated responses when queried about content unrelated to the input audio. To mitigate this, we generated additional instruction data by prompting the LLM to produce questions unrelated to the audio along with appropriate answers. However, assigning a higher weight to this auxiliary dataset during training resulted in reduced speech understanding capability.\\
As LLMs continue to improve, the LLM-as-a-judge approach has become increasingly common for subjective evaluations. However, the reported metrics can vary depending on the LLM used. We observed that the variation was significantly low when running evaluations multiple times with the same LLM. Furthermore, speech-text LLMs have been found to be sensitive to the prompts used for evaluating foundational tasks such as ASR and S2TT. Although we made every effort to use the prompts recommended by each respective model, the necessary information was not available for some tasks or was provided only for the pre-trained variants.

\section*{Ethics Statement}
The proposed SIFT-50M dataset is constructed from three publicly available and widely used speech corpora: Common Voice, MLS, and VCTK. The SIFT-50M dataset is released under the CDLA-Sharing-1.0 license. We acknowledge the potential risks associated with the capability of controllable speech generation, such as misuse for deepfakes, voice impersonation, or the spread of misinformation. Although this task involves inherent complexities, our dataset is curated to limit control to generic, broad parameters. SIFT-LLM only extracts, and we only use non-specific features that cannot be used for identification.

\section*{Acknowledgements}
We would like to thank Leif Rädel, Hayoon Kim, Melanie McHale, Adam Kaplan, Akshat Jaiswal, Subhadeep Maji, Aparna Khare, Anna Piunova, Mohammed Hethnawi, Harish Mallidi, Haolin Chen, Mateusz Łajszczak, Roberto Barra-Chicote, Luyao Wang, Berke Toptas, Sunwoo Kim, Jing Liu, Yang Li,  Ajay Srinivasamurthy, Nathan Susanj, Volker Leutnant, Sri Garimella, and Athanasis Mouchtaris for their support and valuable feedback.

% Bibliography entries for the entire Anthology, followed by custom entries
%\bibliography{anthology,custom}
% Custom bibliography entries only

\onecolumn
\appendix

\section{SIFT-50M Dataset}

\begin{table}[h]
  \centering
  \small
  \renewcommand{\arraystretch}{1.2} % Increase row spacing by 50%
  \begin{tabular}{p{14.0cm}}
    \toprule
    \textbf{Closed-ended: Acoustic-Level} \\
    \textbf{Question}: ¿Hay distorsión en el audio?\\
    \textbf{Answer}: Casi no hay ruido de distorsión en el audio.\\
    \hline 
    \textbf{Closed-ended: Acoustic-Level} \\
    \textbf{Question}: What is the transcript of the audio, and what emotions can be inferred from it?\\
    \textbf{Answer}: The transcript of the audio is `A girl smiles while interacting with medical personnel'. The emotions that can be inferred from the transcript are happiness and cooperation.\\
    \hline 
    \textbf{Closed-ended: Word-Align}       \\
    \textbf{Question}: Wie verändert sich die durchschnittliche Tonhöhe im Laufe des gesprochenen Satzes? \\
    \textbf{Answer}:  Die durchschnittliche Tonhöhe des gesprochenen Satzes beginnt mit einer ziemlich tiefen Tonhöhe, sinkt dann zu einem sehr tiefen Ton und steigt dann allmählich zu einer mittleren Tonhöhe an. Am Ende des Satzes sinkt die Tonhöhe erneut auf eine sehr tiefe Tonhöhe ab. \\
    \hline
    \textbf{Closed-ended: Comparison}        \\
    \textbf{Question}: Qual è la differenza di tonalità tra i due audio? \\
    \textbf{Answer}: Il primo audio ha una tonalità molto alta, mentre il secondo ha una tonalità abbastanza alta. \\
    \hline
    \textbf{Open-ended}  \\
    \textbf{Question}: Dans quelle mesure la r\'everb\'eration et le niveau de distorsion per\c{c}us dans l'audio affectent-ils la clart\'e de la transcription et la compr\'ehension globale du message?  \\
    \textbf{Answer}: La r\'everb\'eration l\'eg\`erement lointain et le niveau de distorsion bruyant rendent la transcription moins claire, compliquant la compr\'ehension pr\'ecise du message. Cela n\'ecessite un effort suppl\'ementaire de l'auditeur pour interpr\'eter le contenu, ce qui peut entra\^{\i}ner des malentendus ou des interpr\'etations erron\'ees.\\
    \hline
    \textbf{Controllable Generation}  \\
    \textbf{Question}: Synthetisieren Sie Sprache mit einer mäßig schnellen Geschwindigkeit, einer mittelmäßigen Lautstärke und einer sehr monotonen Tonhöhenvariation. Der Transkriptionstext lautet: `Auf Kamelen ritten sie durchs Morgenland.' \\
    \textbf{Answer}: \textit{<reference to original audio>} \\
    \bottomrule
  \end{tabular}
  \caption{Representative multilingual examples from SIFT-50M for various instruction categories.}
  \label{tab:qa_types_multilingual}
\vspace{1.0cm}
\end{table}

\begin{figure*}[!hbtp]  % Use figure* for spanning both columns
    \centering
    \includegraphics[width=0.85\textwidth, height=0.5\textheight, keepaspectratio]{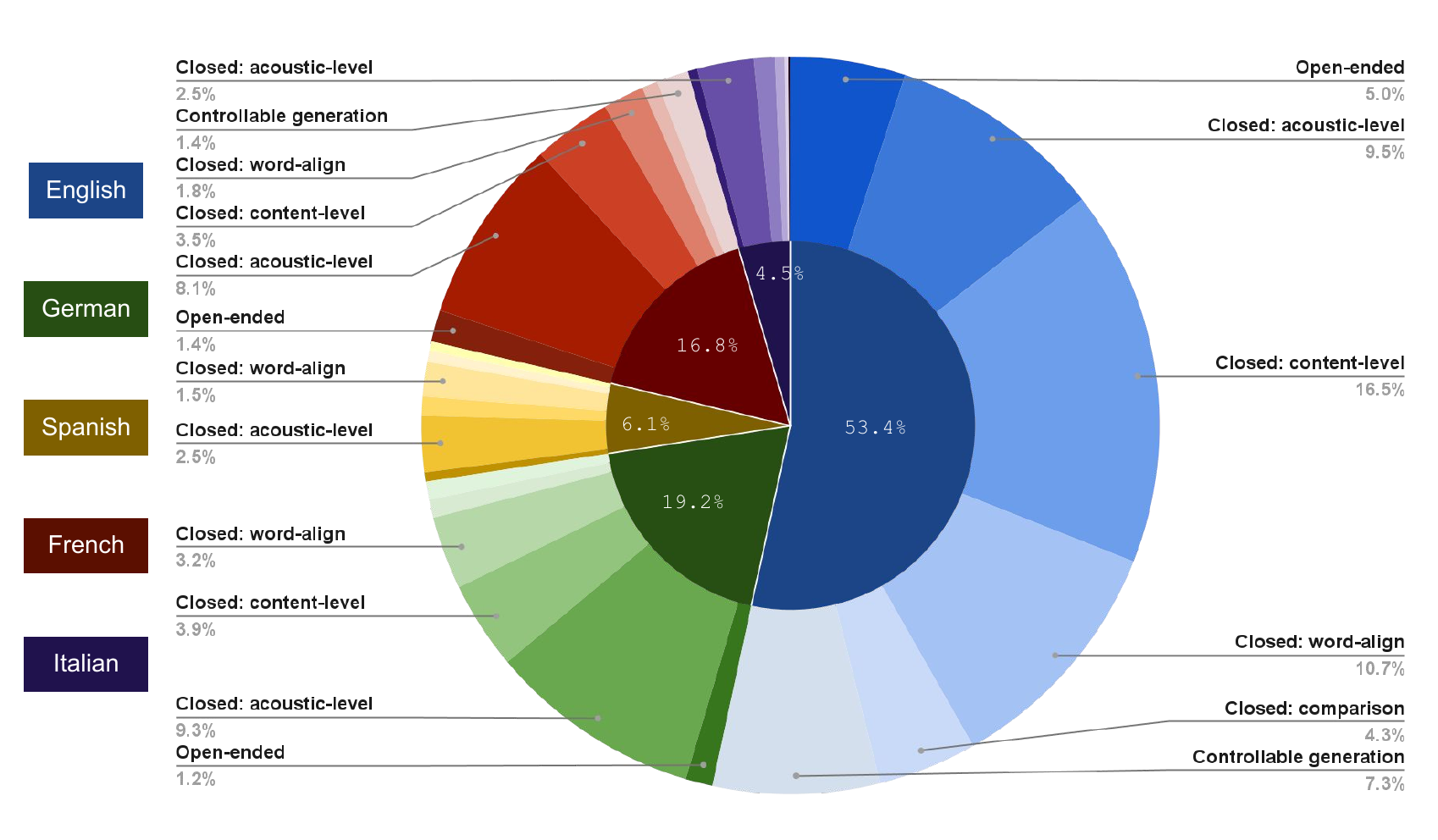}  % Adjust width as needed
    \caption{Dataset distribution showing the multi-lingual nature of SIFT-50M and the different categories within each language. }
    \label{fig:data_distributed_sift}
\end{figure*}

\begin{figure*}[!hbtp]  % Use figure* for spanning both columns
    \centering
    \includegraphics[width=0.7\textwidth, height=0.5\textheight, keepaspectratio]{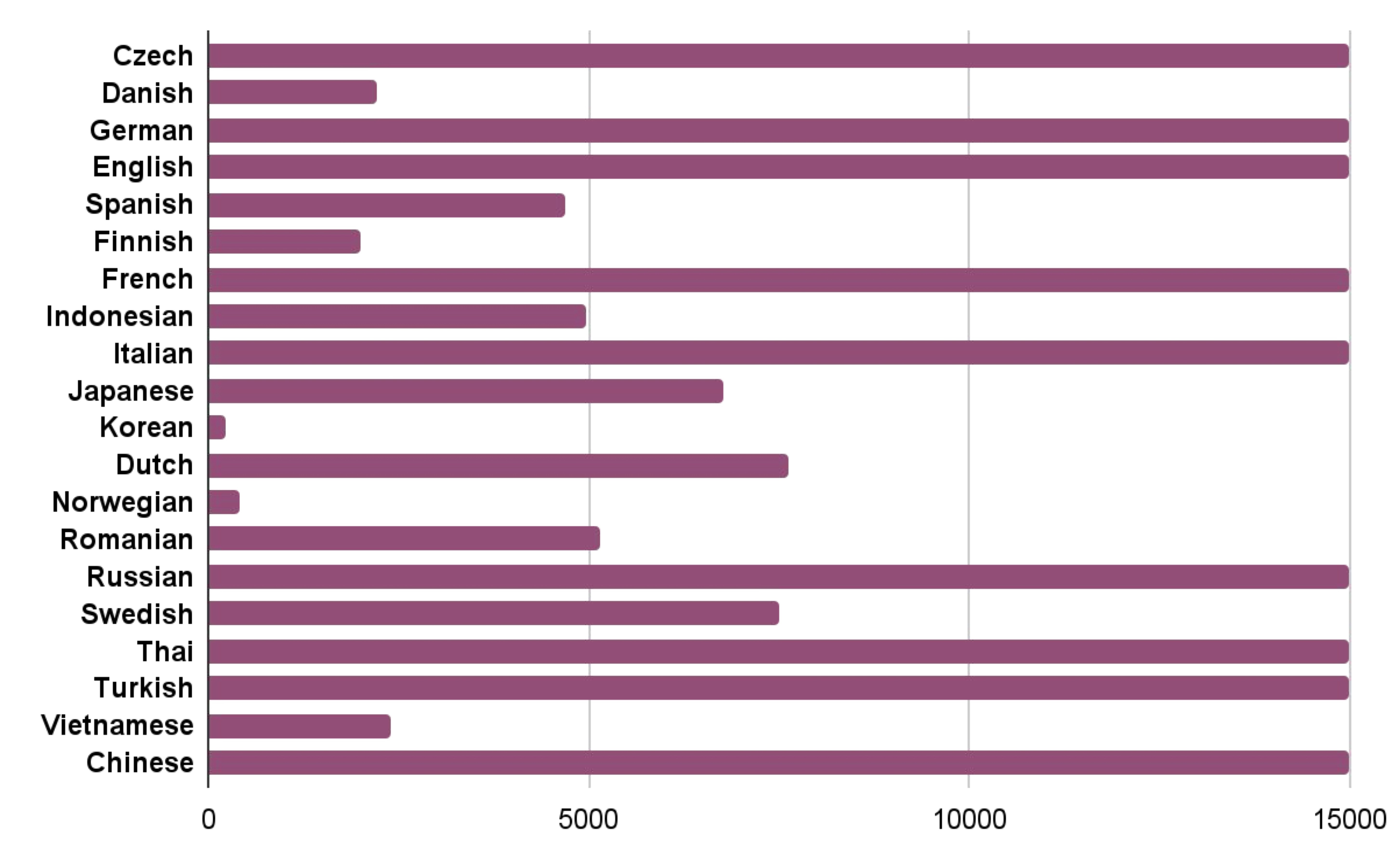}  % Adjust width as needed
    \caption{Distribution of the number of examples per language in the acoustic-based language ID (LID) task that is part of the closed-ended instructions.}
    \label{fig:data_distributed_lang}
\end{figure*}

\newpage

\section{Metadata Extraction}
\subsection{Categorical Values of Acoustic-Level Metadata}
\label{sec:appendix_metadata_map}

\begin{table*}[!hptb]
    \centering
        \resizebox{\textwidth}{!}{
    \begin{tabular}{l l l l l l l l}
        \toprule
        \textbf{Average Pitch} & \textbf{Pitch Variation} & \textbf{Speaking Rate} & \textbf{Average Intensity} & \textbf{Noise Level} & \textbf{Reverberation} & \textbf{Distortion Level} & \textbf{Speech Quality} \\
        \midrule
        Very low pitch & Very monotone & Very slowly & Softly & Very noisy & Very distant-sounding & Extremely noisy & Very bad speech quality \\
        Quite low pitch & Quite monotone & Quite slowly & Moderate volume & Noisy & Distant-sounding & Very noisy & Bad speech quality \\
        Slightly low pitch & Slightly monotone & Slightly slowly & Loudly & Slightly noisy & Slightly distant-sounding & Noisy & Slightly bad speech quality \\
        Moderate pitch & Moderate intonation & Moderate speed &  & Balanced in clarity & Slightly close-sounding & Slightly noisy & Moderate speech quality \\
        Slightly high pitch & Slightly expressive & Slightly fast &  & Slightly clean & Very close-sounding & Almost no noise & Great speech quality \\
        Quite high pitch & Quite expressive & Quite fast &  & Clean &  & Very clear & Wonderful speech quality \\
        Very high pitch & Very expressive & Very fast &  & Very clean &  &  &  \\
        \bottomrule
    \end{tabular}
    }
    \caption{Categorical values of acoustic-level metadata}
    \label{tab:metadata_mapping}
\vspace{-4mm}
\end{table*}

\subsection{Sample Metadata}
We show an example of metadata from the VCTK dataset after mapping to categorical values.

\begin{tcolorbox}[breakable, colframe=blue!40!black, colback=blue!2!white]
\small
\texttt{%
\{\\
\hspace*{2em}\lqquote acoustic\_level\rqquote: \{\\
\hspace*{4em}\lqquote average\_pitch\rqquote: \lqquote slightly low pitch\rqquote,\\
\hspace*{4em}\lqquote pitch\_variation\rqquote: \lqquote quite monotone\rqquote,\\
\hspace*{4em}\lqquote speaking\_rate\rqquote: \lqquote moderate speed\rqquote,\\
\hspace*{4em}\lqquote average\_intensity\rqquote: \lqquote moderate volume\rqquote,\\
\hspace*{4em}\lqquote noise\_level\rqquote: \lqquote slightly noisy\rqquote,\\
\hspace*{4em}\lqquote reverberation\rqquote: \lqquote very close-sounding\rqquote,\\
\hspace*{4em}\lqquote distortion\_level\rqquote: \lqquote almost no noise\rqquote,\\
\hspace*{4em}\lqquote speech\_quality\rqquote: \lqquote moderate speech quality\rqquote\\
\hspace*{2em}\},\\[1mm]
\hspace*{2em}\lqquote content\_level\rqquote: \{\\
\hspace*{4em}\lqquote phonemes\rqquote: \lqquote SH|IY1| |K|AE1|N| |S|K|UW1|P| |DH|IY1|Z| |TH|IH1|NG|Z| ...\rqquote,\\
\hspace*{4em}\lqquote age\_category\rqquote: \lqquote twenties\rqquote,\\
\hspace*{4em}\lqquote accent\rqquote: \lqquote English\rqquote,\\
\hspace*{4em}\lqquote gender\rqquote: \lqquote Female\rqquote,\\
\hspace*{4em}\lqquote transcript\rqquote: \lqquote She can scoop these things into three red bags, and we will go meet her Wednesday at the train station.\rqquote,\\
\hspace*{2em}\}\\
\}
}

\end{tcolorbox}

\subsubsection{Sample Word-level Metadata}

\begin{tcolorbox}[breakable, colframe=blue!40!black, colback=blue!2!white]
\small
\texttt{%
[\\
\hspace*{2em}\{\\
\hspace*{4em}\lqquote word\rqquote: \lqquote nodded\rqquote,\\
\hspace*{4em}\lqquote position\rqquote: 1,\\
\hspace*{4em}\lqquote average\_pitch\rqquote: \lqquote slightly low pitch\rqquote,\\
\hspace*{4em}\lqquote pitch\_variation\rqquote: \lqquote very monotone\rqquote,\\
\hspace*{4em}\lqquote average\_intensity\rqquote: \lqquote moderate volume\rqquote\\
\hspace*{2em}\},\\[1mm]
\hspace*{2em}\{\\
\hspace*{4em}\lqquote word\rqquote: \lqquote his\rqquote,\\
\hspace*{4em}\lqquote position\rqquote: 2,\\
\hspace*{4em}\lqquote average\_pitch\rqquote: \lqquote slightly low pitch\rqquote,\\
\hspace*{4em}\lqquote pitch\_variation\rqquote: \lqquote very monotone\rqquote,\\
\hspace*{4em}\lqquote average\_intensity\rqquote: \lqquote moderate volume\rqquote\\
\hspace*{2em}\},\\[1mm]
\hspace*{2em}\{\\
\hspace*{4em}\lqquote word\rqquote: \lqquote head\rqquote,\\
\hspace*{4em}\lqquote position\rqquote: 3,\\
\hspace*{4em}\lqquote average\_pitch\rqquote: \lqquote quite low pitch\rqquote,\\
\hspace*{4em}\lqquote pitch\_variation\rqquote: \lqquote very monotone\rqquote,\\
\hspace*{4em}\lqquote average\_intensity\rqquote: \lqquote moderate volume\rqquote\\
\hspace*{2em}\}\\
]
}
\end{tcolorbox}

\section{Natural Language Instruction Generation}

\subsection{LLM Hyperparameters}
We provide the LLM hyper-parameters used for generating instructions in Table \ref{tab:llm_hyper_params}. Mixtral 8x7B model was run on a g5.48xlarge\footnote{\url{https://aws.amazon.com/ec2/instance-types/g5/}} instance (8 A10 GPUs), while Amazon Nova Pro was used on AWS Bedrock\footnote{\url{https://aws.amazon.com/bedrock/}}.

\begin{table}[h]
  \centering
  \small
  \renewcommand{\arraystretch}{1.0} % Increase row spacing by 50%
\begin{tabular}{c|cc}
\toprule
\textbf{Parameter} & 
\textbf{Mixtral 8x7B} &
\textbf{Amazon Nova Pro} \\
\midrule
Temperature & 0.7 & 0.7 \\
Top \textit{p} & 0.9 & -  \\
Max. tokens & 2048 & 2048 \\
\bottomrule
\end{tabular}
\caption{LLM hyperparameters used during instruction generation }
\label{tab:llm_hyper_params}
\end{table}

\vspace{-0.2cm}
\subsection{Data Format}
\label{sec:appendix_instruction_format}
% \vspace{0.1cm}
Below is an example of how SIFT-50M data is organized in the ChatML\footnote{\url{https://platform.openai.com/docs/api-reference/chat/create}}/Messages API\footnote{\url{https://docs.anthropic.com/en/api/messages}} format.

\begin{lstlisting}[language=json, linewidth=1.0\textwidth]
{
   (*@\lqquote@*)id(*@\rqquote@*): (*@\lqquote@*)<instance_id>(*@\rqquote@*),
   (*@\lqquote@*)messages(*@\rqquote@*): [
      {
         (*@\lqquote@*)role(*@\rqquote@*): (*@\lqquote@*)user(*@\rqquote@*),
         (*@\lqquote@*)content(*@\rqquote@*): [
            {(*@\lqquote@*)text(*@\rqquote@*): (*@\lqquote@*)Which audio has a speaker with a Scottish accent?(*@\rqquote@*)},
            {(*@\lqquote@*)audio_path(*@\rqquote@*): (*@\lqquote@*)<audio_path_1>(*@\rqquote@*)},
            {(*@\lqquote@*)audio_path(*@\rqquote@*): (*@\lqquote@*)<audio_path_2>(*@\rqquote@*)}
         ]
      },
      {
         (*@\lqquote@*)role(*@\rqquote@*): (*@\lqquote@*)assistant(*@\rqquote@*),
         (*@\lqquote@*)content(*@\rqquote@*): [{(*@\lqquote@*)text(*@\rqquote@*): (*@\lqquote@*)The second audio has a Scottish-accented speaker.(*@\rqquote@*)}]
      }
   ]
}
\end{lstlisting}

\section{Training Setup}
\label{appendix:training_setup}

\subsection{Pre-training Datasets}
\begin{table}[h]
  \centering
  \small
  \renewcommand{\arraystretch}{1.15} % Increase row spacing by 50%
  \begin{tabular}{p{4.2cm}cp{5.7cm}c}
    \toprule
    % \textbf{\begin{minipage}[t]{4cm}\centering Task\end{minipage}} & \textbf{Dataset} & \textbf{\#Hours} & \textbf{\#Samples} \\
    \textbf{\begin{minipage}[t]{4cm}\centering Task\end{minipage}} & \textbf{Task Wt.} & \textbf{Dataset} & \textbf{\#Hours}\\
    \midrule
    \multirow{4}{4cm}{\begin{minipage}[t]{4cm}\centering Automatic Speech Recognition (ASR)\end{minipage}} &
    \multirow{4}{*}{0.35} & Librispeech \cite{panayotov2015librispeech} & 960 \\
                         & & MLS \cite{pratap2020mls} & 50K \\
                         & & FLEURS \cite{conneau2023fleurs} & 1.4K \\
                         & & People's Speech \cite{galvez2021people} & 30K \\
    \hline
    \begin{minipage}[t]{5cm}\centering Emotion Recognition (ER)\end{minipage} & 0.05 & MSP-Podcast \cite{lotfian2017building} & 237 \\
    \hline
    \multirow{3}{4cm}{\begin{minipage}[t]{4.0cm}\centering Speaker-Attributed ASR (SA-ASR)\end{minipage}} &
    \multirow{3}{*}{0.05} & AMI Meeting Corpus \cite{carletta2005ami} & 100 \\
                         & & ICSI Meeting Corpus \cite{janin2003icsi} & 70 \\
                         & & Fisher \cite{cieri2004fisher} & 2K \\
    \hline
    \begin{minipage}[t]{4cm}\centering Intent Classification (IC)\end{minipage} & \multirow{2}{*}{0.05} & SLURP \cite{bastianelli2020slurp}  & 58 \\
    \begin{minipage}[t]{4cm}\centering Slot Entity Recognition (SER)\end{minipage} & & SLURP \cite{bastianelli2020slurp} & 58 \\
    \hline
    \multirow{4}{4cm}{\begin{minipage}[t]{4cm}\centering Speech Translation \hspace{1.5cm} (S2TT/ S2ST)\end{minipage}} &
    \multirow{4}{*}{0.20} & CoVoST \cite{wang2021covost} & 2.9K \\
                         & & CVSS \cite{jia2022cvss} & 1.1K \\
                         & & VoxPopuli \cite{wang2021voxpopuli} & 1.8K \\
                         & & FLEURS \cite{conneau2023fleurs} & 1.4K \\
    \hline
    \multirow{3}{4cm}{\begin{minipage}[t]{4cm}\centering Text to Speech (TTS)\end{minipage}} &
    \multirow{3}{*}{0.30} & MLS \cite{pratap2020mls} & 50K \\
                         & & FLEURS \cite{conneau2023fleurs} & 1.4K \\
                         & & People's Speech \cite{galvez2021people} & 30K \\
    \bottomrule
  \end{tabular}
  \caption{Details of the datasets used during the pre-training stage and the weights assigned to different tasks.}
  \label{tab:pretraining}
\end{table}

\subsection{Training Hyper-parameters}
We provide the training hyper-parameters used for SIFT-LLM in Table~\ref{tab:hyper_params}. All models were trained on four p4d.24xlarge\footnote{\url{https://aws.amazon.com/ec2/instance-types/p4/}} instances (32 A100 GPUs in total, each with 40GB of memory), with training for 200K steps taking approximately 3 days.

\begin{table}[h]
  \centering
  \small
  \renewcommand{\arraystretch}{1.1} % Increase row spacing by 50%
\begin{tabular}{c|cc|cc}
\toprule
\textbf{Parameter} & \textbf{PT: Stage 1} & \textbf{IFT: SIFT-LLM} & \textbf{PT: Stage 2} & \textbf{IFT: SIFT-LLM GEN} \\
\midrule
Batch Size & 256 & 128 & 256 & 128 \\
\# Steps & 200K & 200K & 200K & 200K \\
\# Warmup Steps & 1000 & 500 & 500 & 500 \\
Max Learning Rate & 1e-4 & 1e-4 & 1e-4 & 1e-4 \\
LoRA rank & 16 & 16 & 128 & 128 \\
LoRA $\alpha$ & 8 & 8 & 64 & 64 \\
\# Trainable Params & 8.7M & 8.7M & 590M (HuBERT) & 590M (HuBERT) \\
\bottomrule
\end{tabular}
\caption{Training hyperparameters used during pre-training (PT) and instruction fine-tuning (IFT).}
\label{tab:hyper_params}
\end{table}

% \newpage

\subsection{Pre-Training Prompt Template Examples}

Below are examples of instruction templates used during pre-training:

\begin{tcolorbox}[breakable, colframe=blue!40!black, colback=blue!2!white]
\textbf{ASR:} \\
\hspace*{0.8cm}Transcribe what the speaker is saying.\\
\hspace*{0.8cm}Decode this \$language speech.\\
\\
\textbf{ER:} \\
\hspace*{0.8cm}Get the most probable emotion in the utterance.\\
\hspace*{0.8cm}Annotate the speech and obtain the most prominent emotion.\\
\\
\textbf{SA-ASR:} \\
\hspace*{0.8cm}Transcribe what each of the speakers says and assign speaker labels to transcribed segments. Format the output as text interspersed with speaker tags, where each utterance is followed by its speaker in JSON format (e.g., {"speaker": "spk\_id"}). Use integer speaker ids starting with 0.\\
\\
\textbf{IC:} \\
\hspace*{0.8cm}From this audio, what intent does the speaker have? Options include: joke, definition, takeaway\_query, and social\_query.\\
\\
\textbf{SER:} \\
\hspace*{0.8cm}Can you identify the entity that fits the device\_type slot?\\
\\
\textbf{S2TT:} \\
\hspace*{0.8cm}Recognize and then translate this utterance to text in \$target\_language.\\
\hspace*{0.8cm}Translate text from \$source\_language to \$target\_language for this audio.\\
\\
\textbf{S2ST:} \\
\hspace*{0.8cm}Can you translate this recording to \$target\_language\\
\\
\textbf{TTS:} \\
\hspace*{0.8cm}Can you convert the text to speech: \$transcription?
\end{tcolorbox}

\section{Evaluation Setup}
\label{appendix:detailed_results}

\subsection{Instruction-Following Benchmarks}

\begin{table}[h]
  \centering
  \small
  \renewcommand{\arraystretch}{1.1} % Increase row spacing by 50%
\begin{tabular}{lcc}
\toprule
\textbf{Test Set} & \textbf{\# Samples} & \textbf{Languages} \\
\midrule
Dynamic-Superb \cite{huang2024dynamic2} & 10,400 & English \\
AIR-Bench Chat \cite{yang2024air} & 793  & English \\
EvalSIFT (Closed-Ended) & 10K (2K per lang)  & English, German, French, Italian, Spanish \\
EvalSIFT (Open-Ended) & 10K (2K per lang)  & English, German, French, Italian, Spanish \\
EvalSIFT (Control. Generation) & 10K (2K per lang)  & English, German, French, Italian, Spanish \\
\bottomrule
\end{tabular}
\caption{Details of benchmarks used for instruction-following evaluation.}
\label{tab:if_benchmarks}
\end{table}

\subsection{Foundational Tasks Benchmarks}

\begin{table}[h]
  \centering
  \small
  \renewcommand{\arraystretch}{1.1} % Increase row spacing by 50%
\begin{tabular}{lcc}
\toprule
\textbf{Task} & \textbf{Test Set} & \textbf{\# Samples}\\
\midrule
\multirow{4}{*}{ASR} & Librispeech Test-Clean \cite{panayotov2015librispeech} & 2,620\\
 & Librispeech Test-Other \cite{panayotov2015librispeech} & 2,560\\
 & People's Speech \cite{galvez2021people} & 34,898\\
 & FLEURS (en, de, fr, it, es) \cite{conneau2023fleurs} & 350 * 5 = 1,750 \\
\hline
\multirow{2}{*}{ER} & MSP-test1 \cite{lotfian2017building} & 30,647 \\
 & MSP-test2 \cite{lotfian2017building} & 14,815 \\
\hline
IC & SLURP \cite{bastianelli2020slurp} & 13,078\\
SER & SLURP \cite{bastianelli2020slurp} & 11,585\\
\hline
\multirow{4}{*}{S2TT} & CoVost2 (de $\rightarrow$ en) \cite{wang2021covost} & 13,509 \\
 & CoVost2 (fr $\rightarrow$ en) \cite{wang2021covost} & 14,760 \\
 & CoVost2 (it $\rightarrow$ en) \cite{wang2021covost} & 8,945 \\
 & CoVost2 (es $\rightarrow$ en) \cite{wang2021covost} & 13,221\\
\bottomrule
\end{tabular}
\caption{Details of benchmarks used for foundational tasks evaluation.}
\label{tab:foundation_benchmarks}
\end{table}

\subsection{Evaluation Prompt Templates}

As model performance can be sensitive to the prompts used \cite{wang2024audiobench}, we use the prompts recommended by the corresponding model wherever we can find for evaluation of SALMONN-7B\footnote{\url{https://github.com/bytedance/SALMONN/blob/main/prompts/test_prompt.json}}, Qwen2-Audio-Instruct\footnote{\url{https://github.com/QwenLM/Qwen2-Audio/blob/main/eval_audio/EVALUATION.md}} and O-ASQA-LLM\footnote{\url{https://github.com/YuanGongND/ltu/blob/main/README.md\#for-ltu-as-openasqa}}.

\begin{table}[h]
  \centering
  \small
  \renewcommand{\arraystretch}{1.1} % Increase row spacing by 50%
\begin{tabular}{lcp{10.5cm}}
\toprule
\textbf{Task} & \textbf{Model} & \textbf{Prompt Template}\\
\midrule
\multirow{10}{*}{ASR} & \multirow{5}{*}{SALMONN-7B} & [en] Recognize the speech and give me the transcription.\\
 & & [de] Hören Sie sich die Rede an und schreiben Sie ihren Inhalt auf.\\
 & & [fr] Écoutez le discours et écrivez son contenu.\\
 & & [it] Ascolta il discorso e scrivi il suo contenuto.\\
 & & [es] Escuche el discurso y escriba su contenido.\\
 \cline{2-3}
 & Qwen2-Audio-Instruct & [en/de/fr/it/es] Recognize the speech in \$language:\\
 \cline{2-3}
 & \multirow{2}{*}{O-ASQA-LLM} & [en] Can you identify the spoken text?\\
 & & [de/fr/it/es] Can you identify the spoken text?  Your output MUST be in \$language.\\
 \cline{2-3}
 & \multirow{2}{*}{ SIFT-LLM} & [en] Transcribe what the speaker says.\\
 & & [de/fr/it/es] Transcribe this \$language speech.\\
\hline
\multirow{4}{*}{ER} & SALMONN-7B & Describe the emotion of the speaker in one word.\\
 & Qwen2-Audio-Instruct & Recognize the emotion with keywords in English:\\
 & O-ASQA-LLM & Identify the most likely emotion in the following speech.\\
 & SIFT-LLM & Identify the most likely emotion in the following speech.\\
\hline
\multirow{4}{*}{S2TT} & SALMONN-7B & Listen to the speech and translate it into \$target\_language.\\
 & Qwen2-Audio-Instruct & Translate the speech into \$target\_language:\\
 & O-ASQA-LLM & Translate the audio to \$target\_language, returning only the translated text.\\
 & SIFT-LLM & Translate the audio to \$target\_language, returning only the translated text.\\
\bottomrule
\end{tabular}
\caption{Prompts used to evaluate different speech-text LLMs on foundational tasks.}
\label{tab:eval_prompts}
\end{table}

\section{Evaluation Results}
\label{appendix:detailed_results}
\subsection{EvalSIFT Closed-Ended Results}

\begin{table}[h]
  \centering
  \small
  \renewcommand{\arraystretch}{1.1} % Increase row spacing by 50%
\begin{tabular}{lcccc}
\toprule
\textbf{Model} & \textbf{Acoustic-Level} & \textbf{Content-Level} & \textbf{Word-Align} & \textbf{Comparison} \\
\midrule
SALMONN-7B & 16.1 & 34.6 & 9.7 & 27.3 \\
Qwen2-Audio-Instruct & 19.1 & 37.5 & 16.8 & 27.3 \\
O-ASQA-LLM & 12.8 & 39.6 & 16.7 & 22.7 \\
SIFT-LLM & 58.4 & 41.9 & 28.6 & 55.5 \\
\bottomrule
\end{tabular}
\caption{Breakdown of evaluation results, reported as accuracy (\%) for different sub-categories of the closed-ended set in EvalSIFT (English).}
\label{tab:evalsift_closed}
\end{table}

\subsection{Dynamic-Superb Detailed Results}

{
\scriptsize  % Start smaller font size
\begin{longtable}{p{6.9cm}cccc}
\label{tab:longtable_example}\\
\toprule
\textbf{Task} & \textbf{SALMONN} & \textbf{Qwen2-Audio-Instruct} & \textbf{O-ASQA-LLM} & \textbf{SIFT-LLM} \\
\midrule
\endfirsthead

\toprule
\textbf{Task} & \textbf{SALMONN-7B} & \textbf{Qwen2-Audio-Inst.} & \textbf{O-ASQA-LLM} & \textbf{SIFT-LLM} \\
\midrule
\endhead

\bottomrule
\endfoot

%--------------- BEGIN TABLE ROWS ---------------%
AccentClassification\_AccentdbExtended & 13 & 14.5 & 17 & 50.5 \\
DialogueEmotionClassification\_DailyTalk & 16.5 & 46 & 57.29 & 65.83 \\
EmotionRecognition\_MultimodalEmotionlinesDataset & 32.32 & 38.5 & 51 & 60.41 \\
HowFarAreYou\_3DSpeaker & 23.12 & 31 & 28 & 28.5 \\
SpoofDetection\_ASVspoof2015 & 61.5 & 20.1 & 14.07 & 5.5 \\
SpoofDetection\_ASVspoof2017 & 65 & 27.64 & 23.62 & 23 \\
StressDetection\_MIRSD & 2 & 24.5 & 18.59 & 28.5 \\
\midrule
\textbf{Paralinguistics} & \textbf{30.49} & \textbf{28.89} & \textbf{29.94} & \textbf{37.46} \\
\midrule
BirdSoundDetection\_Warblrb10k & 74 & 76.88 & 40.91 & 35.35 \\
ChordClassification\_AcousticGuitarAndPiano & 7.5 & 40.2 & 49.5 & 46.5 \\
EnvironmentalSoundClassification\_ESC50-Animals & 42 & 62.31 & 10.61 & 42.71 \\
EnvironmentalSoundClassification\_ESC50-ExteriorAndUrbanNoises & 10 & 60.1 & 14 & 36.36 \\
EnvironmentalSoundClassification\_ESC50-HumanAndNonSpeechSounds & 54.04 & 47.47 & 24.5 & 43.5 \\
EnvironmentalSoundClassification\_ESC50-InteriorAndDomesticSounds & 5.56 & 31.5 & 34.5 & 48.72 \\
EnvironmentalSoundClassification\_ESC50-NaturalSoundscapesAndWaterSounds & 28.5 & 55.84 & 25.5 & 46.23 \\
\midrule
\textbf{Audio} & \textbf{31.66} & \textbf{53.47} & \textbf{28.5} & \textbf{42.77} \\
DialogueActClassification\_DailyTalk & 42.35 & 33.17 & 23.62 & 40.7 \\
DialogueActPairing\_DailyTalk & 53.77 & 40.1 & 41.41 & 47.24 \\
SarcasmDetection\_Mustard & 46.5 & 47.5 & 50.75 & 66 \\
\midrule
\textbf{Semantics} & \textbf{47.54} & \textbf{40.26} & \textbf{38.59} & \textbf{51.31} \\
\midrule
EnhancementDetection\_LibriTTS-TestClean\_WHAM & 28.93 & 52.26 & 48.5 & 82.5 \\
NoiseDetection\_LJSpeech\_MUSAN-Gaussian & 49 & 49.5 & 52.5 & 72.5 \\
NoiseDetection\_LJSpeech\_MUSAN-Music & 49.75 & 46 & 51.28 & 86.93 \\
NoiseDetection\_LJSpeech\_MUSAN-Noise & 50 & 49 & 52.31 & 86.5 \\
NoiseDetection\_LJSpeech\_MUSAN-Speech & 43 & 46.5 & 52.55 & 87.94 \\
NoiseDetection\_VCTK-MUSAN-Gaussian & 57 & 54.5 & 49 & 74 \\
NoiseDetection\_VCTK\_MUSAN-Music & 47.72 & 54.5 & 51.52 & 84.92 \\
NoiseDetection\_VCTK\_MUSAN-Noise & 45.5 & 58.29 & 47.98 & 89 \\
NoiseDetection\_VCTK\_MUSAN-Speech & 36.5 & 55.5 & 52.28 & 87.5 \\
NoiseSNRLevelPrediction\_VCTK\_MUSAN-Gaussian & 25.13 & 22.8 & 26.63 & 21.11 \\
NoiseSNRLevelPrediction\_VCTK\_MUSAN-Music & 26.74 & 16.67 & 30.15 & 25.5 \\
NoiseSNRLevelPrediction\_VCTK\_MUSAN-Noise & 29.29 & 18.18 & 28.28 & 23 \\
NoiseSNRLevelPrediction\_VCTK\_MUSAN-Speech & 23.5 & 28.21 & 28.43 & 25.13 \\
ReverberationDetection\_LJSpeech\_RirsNoises-LargeRoom & 9 & 47.5 & 45.5 & 65.5 \\
ReverberationDetection\_LJSpeech\_RirsNoises-MediumRoom & 8.54 & 48 & 51.5 & 54.82 \\
ReverberationDetection\_LJSpeech\_RirsNoises-SmallRoom & 3.5 & 48 & 50.51 & 57.79 \\
ReverberationDetection\_VCTK\_RirsNoises-LargeRoom & 12.5 & 46 & 54.5 & 67.84 \\
ReverberationDetection\_VCTK\_RirsNoises-MediumRoom & 15 & 46 & 48.24 & 60 \\
ReverberationDetection\_VCTK\_RirsNoises-SmallRoom & 10 & 46 & 50.51 & 56 \\
\midrule
\textbf{Degradation} & \textbf{30.03} & \textbf{43.86} & \textbf{45.9} & \textbf{63.6} \\
\midrule
LanguageIdentification\_VoxForge & 31.5 & 89.95 & 50 & 88.5 \\
SpeechCommandRecognition\_GoogleSpeechCommandsV1 & 18.88 & 67.19 & 70.56 & 64.32 \\
SpeechDetection\_LJSpeech & 57.87 & 40.7 & 67 & 62.28 \\
SpeechDetection\_LibriSpeech-TestClean & 45.73 & 53.77 & 49.49 & 49.95 \\
SpeechDetection\_LibriSpeech-TestOther & 46.5 & 48.5 & 50.75 & 45.73 \\
SpeechTextMatching\_LJSpeech & 51.26 & 80.2 & 88.72 & 78.28 \\
SpeechTextMatching\_LibriSpeech-TestClean & 52.26 & 86.36 & 83.94 & 72.96 \\
SpeechTextMatching\_LibriSpeech-TestOther & 52.53 & 83.33 & 84.9 & 76.77 \\
SpokenTermDetection\_LJSpeech & 51.27 & 81.87 & 87.05 & 97.97 \\
SpokenTermDetection\_LibriSpeech-TestClean & 47.21 & 73.22 & 81.44 & 97.85 \\
SpokenTermDetection\_LibriSpeech-TestOther & 41.75 & 71.2 & 81.35 & 97.3 \\
\midrule
\textbf{Content} & \textbf{45.16} & \textbf{70.57} & \textbf{72.29} & \textbf{75.63} \\
\midrule
MultiSpeakerDetection\_LibriSpeech-TestClean & 13.2 & 44.95 & 47.24 & 63.13 \\
MultiSpeakerDetection\_VCTK & 39.39 & 45.41 & 47.74 & 73.5 \\
SpeakerCounting\_LibriTTS-TestClean & 9.18 & 24.37 & 21 & 17.17 \\
SpeakerVerification\_LibriSpeech-TestClean & 44.72 & 51.5 & 41.71 & 46.73 \\
SpeakerVerification\_VCTK & 52.76 & 51.5 & 45.73 & 38.19 \\
\midrule
\textbf{Speaker} & \textbf{31.85} & \textbf{43.55} & \textbf{40.68} & \textbf{47.74} \\
%--------------- END TABLE ROWS ---------------%
\end{longtable}
\vspace{-0.4cm}
}

\subsection{Limitations of HuBERT codes}\label{appendix:controllable_results}
To validate the efficacy of discretized HuBERT codes in preserving speech characteristics, we conducted an experiment in which we synthesized a waveform from the HuBERT codes extracted from an original audio waveform. We refer to the resulting audio as re-synthesized audio. In this experiment, we used a randomly selected reference speaker signal in the vocoder. Using the same metrics as in Table~\ref{tab:controllable_eval}, we compared the speech characteristics of the re-synthesized audio with those of the original audio. Ideally, the re-synthesized audio should match the characteristics of the original audio. However, the results, shown in Table~\ref{tab:controllable_eval_resynth_ref}, indicate that while speaking rate exhibited some similarity, other characteristics differed considerably.

In a further experiment, we assessed the effect of the reference speaker signal used in the vocoder. First, we selected original audio samples from the source speech corpus, whose metadata was used to construct a given controllable generation instruction. We then compared two outputs: (1) the audio re-synthesized from HuBERT codes extracted from the original audio, and (2) the audio generated by SIFT-LLM GEN for the corresponding instruction. For each example, the same reference speaker signal was used in the vocoder. The results, reported in Table~\ref{tab:controllable_eval_resynth_hyp}, show that pitch variation and intensity exhibit a much stronger correlation when the same reference speaker is used. We also observed that gender predictions matched approximately 96\% between the SIFT-LLM GEN generated audio and the re-synthesized audio, whereas the match was only around 50\% when compared against the gender specified in the controllable instruction. We hypothesize this limitation is due to incomplete representation of speech in HuBERT codes—i.e, they primarily capture semantic content and limited stationary aspects such as speaking rate, pitch, and intensity.

\begin{table}[h]
  \centering
  \small
  \renewcommand{\arraystretch}{1.0} % Increase row spacing by 50%
  \begin{tabular}{lccc}
    \toprule
    \textbf{Metric} & \textbf{Pitch variation} & \textbf{Speaking rate}& \textbf{Intensity}\\
    \midrule
    MAE ($\downarrow$)    & 0.99 & 0.24 & 0.32 \\
   QWK ($\uparrow$) & 0.11  & 0.72 & 0.06 \\
   
   $\rho$ ($\uparrow$) & 0.12 & 0.79 & 0.09\\
   \bottomrule
  \end{tabular}
  \caption{Evaluation results on controllable generation, comparing the speech characteristics of resynthesized audio and original audio. }

  \label{tab:controllable_eval_resynth_ref}
\vspace{-2mm}
\end{table}

\begin{table}[h]
  \centering
  \small
  \renewcommand{\arraystretch}{1.0} % Increase row spacing by 50%
  \begin{tabular}{lccc}
    \toprule
    \textbf{Metric} & \textbf{Pitch variation} & \textbf{Speaking rate}& \textbf{Intensity}\\
    \midrule
    MAE ($\downarrow$)    & 0.43 & 0.95 & 0.06 \\
   QWK ($\uparrow$) & 0.62  & 0.23 & 0.10 \\
   
   $\rho$ ($\uparrow$) & 0.60 & 0.48 & 0.15\\
   \bottomrule
  \end{tabular}
  \caption{Evaluation results on controllable generation, comparing the speech characteristics of audio generated by SIFT-LLM GEN and resynthesized audio. }

\label{tab:controllable_eval_resynth_hyp}
\vspace{-2mm}
\end{table}

\section{Prompt Templates}

\subsection{Prompts used for Data Generation}
\label{sec:appendix_data_prompts}
\begin{tcolorbox}[breakable, colframe=blue!40!black, colback=blue!2!white, title=Acoustic/ Content level]
Based on the following audio clip, generate \${noques} different types of questions and corresponding step-by-step answers. 
Questions should be about the audio (eg. asking about its properties). The question should be framed as if someone has uploaded an 
audio (and not its metadata) and asking question about it. It should be answered based on metadata values but answer should not 
mention about metadata being available or so. It should reflect as if it is answered after listening to the audio.
The more complex and diverse the question, the better.
Format each QA pair in a single line as a JSON dictionary (key “q” for question, and “a” for answer, wrapped with \{ and \}). 
Do not include any other explanation.

Metadata:
\${metadata}

\end{tcolorbox}

\begin{tcolorbox}[breakable, colframe=blue!40!black, colback=blue!2!white, title=Word-Align]
Based on the following audio clip, generate \${noques} different types of questions and corresponding step-by-step answers in \${language} language. 
The metadata is provided in \${language} language.  
Questions should be about the audio (eg. asking about its properties). The question should be framed as if someone has uploaded an 
audio (and not its metadata) and asking question about it. It should be answered based on metadata values but answer should not 
mention about metadata being available or so. It should reflect as if it is answered after listening to the audio.
The more complex and diverse the question, the better. The metadata is a list of word level characteristics like position and possible pitch and intensity categories.
Questions can be framed on highlighting considerable changes in pitch and intensity at phrase level, each phrase is combination of consecutive words, determined by position info for each word.
Format each QA pair in a single line as a JSON dictionary (key “q” for question, and “a” for answer, wrapped with \{ and \}). 
Do not include any other explanation.

Metadata:
\${metadata}

\end{tcolorbox}

\begin{tcolorbox}[breakable, colframe=blue!40!black, colback=blue!2!white, title=Controllable Generation]

    You will be provided with metadata describing an audio sample of a person's speech.  
Your task is to generate a prompt in \${language} language that instructs a model to synthesize speech based on a description derived from the given metadata.  

Requirements: \\
- The generated prompt must be grammatically correct, concise, and easy to understand. \\
- It should not reference the transcription, which is provided separately. \\
- Aim for complexity and diversity by selecting random speech characteristics to introduce variability.  
- Do not add details beyond what is provided in the metadata.  

Return only the generated prompt and do not include any other explanation.

Metadata:  
\${metadata}  
\end{tcolorbox}

\subsection{Prompts used for Evaluation}
\label{sec:appendix_eval_prompts}

\begin{tcolorbox}[breakable, colframe=blue!40!black, colback=blue!2!white, title=Closed-ended Evaluation]

Below is a question and the correct/reference answer about an audio:\\
\{
    "question": "\${question}",
    "answer": "\${answer}"
\}

Rate the following answer as correct or not.  If the above answer is correct, "score" should be 1, else 0. Answer: \${hyp}\\
Your output should be a JSON dictionary with keys as "reason" and "score". Do not include preambles, or any additional content.  
\end{tcolorbox}

\begin{tcolorbox}[breakable, colframe=blue!40!black, colback=blue!2!white, title=Open-ended Evaluation]

Below is a question and the correct/reference answer about an audio:\\
\{
    "question": "\${question}",
    "answer": "\${answer}"
\}

Rate the following answer on a scale of 0 to 10. Answer: "\${hyp}"\\
Your output should be a JSON dictionary with keys as "reason" and "score". Do not include preambles, or any additional content. 
\end{tcolorbox}

\section{Examples of SIFT-LLM's reasoning abilities}
\label{sec:reasoning}

\begin{tcolorbox}[breakable, colframe=blue!40!black, colback=blue!2!white]

\textbf{Question 1:} What might be the possible scenario in which this speech was delivered?\\
(Audio: We want to continue to be on the periphery of the European infrastructure system, besides the sins of our people.)\\
\textbf{Reference Answer:} One possible scenario in which this speech could have been delivered is during a conference or meeting where the speaker is discussing their organization's stance on the European infrastructure system.\\
\textbf{Model Response:} Given the formal tone, international accent, and specific vocabulary related to infrastructure and Europe, this speech could have been delivered in an academic or professional setting, possibly at a conference or seminar.\\
\\
\textbf{Question 2:} What is the speaker's concern in the speech?\\
(Audio: How do I strengthen it all in a acceptable form that can be done in two or two and a half hours?)\\
\textbf{Reference Answer:} The speaker's concern is to figure out a way to achieve the task within the given time frame.\\
\textbf{Model Response:} The speaker is concerned about the effort and time required to achieve a certain form or appearance. Specifically, they are questioning how someone could accomplish this in two or two-and-a-half hours. \\
\\
\textbf{Question 3:} Evaluate the effectiveness of the audio for use in a professional setting and propose modifications to enhance its suitability.\\
\textbf{Reference Answer:} The audio is ineffective for a professional setting due to its poor speech quality and slow delivery. Modifications such as re-recording with better equipment, increasing speaking rate, and reducing background noise would enhance its suitability.\\
\textbf{Model Response:} The audio is not suitable for a professional setting due to its very bad speech quality, noisy background, and distant-sounding reverberation. To enhance its suitability, the speaker should record in a quiet environment with minimal background noise. Using a high-quality microphone and ensuring proper distance from the microphone can reduce reverberations. Additionally, post-production editing could be used to clean up the audio and improve clarity.
\end{tcolorbox}

\end{document}